\documentclass[letterpaper, journal,onecolumn,12pt,draftcls]{IEEEtran}

\normalsize

\usepackage[dvips]{graphicx}
\usepackage{graphicx}
\usepackage{caption}
\usepackage{subcaption}

\usepackage{enumitem}
\usepackage{amssymb}
\usepackage{amsmath}

\newtheorem{theorem}{Theorem}
\newtheorem{remark}{Remark}
\newtheorem{lemma}{Lemma}

\begin{document}

\title{Achievable Rate Regions for Two-Way Relay Channel using Nested Lattice Coding}
\author{Sinda Smirani, Mohamed Kamoun, Mireille Sarkiss, Abdellatif Zaidi and Pierre Duhamel%
\thanks{Sinda Smirani, Mohamed Kamoun and Mireille Sarkiss are with Communicating Systems Laboratory, CEA, LIST, BC 94, 91191 Gif Sur Yvette, France. Email: \{sinda.smirani, mohamed.kamoun, mireille.sarkiss\}@cea.fr}
\thanks{Abdellatif Zaidi is with Universit\'e Paris-Est Marne La Vall\'ee, LIGM, 77454 Marne la Vall\'ee Cedex 2, France. Email: abdellatif.zaidi@univ-mlv.fr}%
\thanks{Pierre Duhamel is with CNRS/LSS, Supélec, 91192 Gif-Sur-Yvette, France. E-mail: pierre.duhamel@lss.supelec.fr}}%

\maketitle 

\section*{Abstract}
This paper studies Gaussian Two-Way Relay Channel where two communication nodes exchange messages with each other via a relay. 
It is assumed that all nodes operate in half duplex mode without any direct link between the communication nodes.
A compress-and-forward relaying strategy using nested lattice codes is first proposed.
Then, the proposed scheme is improved by performing a layered coding 
: a common layer is decoded by both receivers and a refinement layer is recovered only by the receiver which has the best channel conditions. 
The achievable rates of the new scheme are characterized and are shown to be higher than those provided by the decode-and-forward strategy in some regions.
\begin{IEEEkeywords}
Compress-and-forward, Gaussian channel, lattice codes, physical-layer network coding, side information, two-way relay channel.
\end{IEEEkeywords}
\section{Introduction}
Consider the Two-Way Relay Channel (TWRC)
that is shown in Fig. \ref{Fig1}. 
Two wireless terminals T$_1$ and T$_2$, with no direct link between them, exchange individual messages 
via a relay. Recently, the capacity characterization of this channel has attracted a lot of interest since TWRC is encountered in various wireless communication scenarios, such as ad-hoc networks, or range extension for cellular and local networks.

While network level routing is the standard option to solve this problem, it has been shown that network coding (NC) strategies provide better
performance by leveraging the side information that is available at each
node. In fact, NC \cite{ieeeIT.acly00} offers rate improvements by combining raw bits or
packets at network layer. The rate performance of the system can be further improved if NC
takes place at the physical layer. 
In this situation, the linear superposition property of the wireless channel is considered as a "code" and can be exploited appropriately to turn interference into a useful signal \cite{mobicom.zll06}. 
In this context, we consider a physical-layer network coding (PNC) architecture in which the overall communication requires two phases, namely a Multiple Access (MAC) phase in which the terminals simultaneously send their messages to the relay and a Broadcast (BC) phase in which the relay transmits a message that is a function of the signals received in the MAC phase. An outer bound on the capacity region of this model is given in \cite{gdrisis.k07, ieeeIT.kmt08}.

Several coding strategies have been proposed for PNC by extending classical relaying strategies such as Amplify-and-Forward (AF), Decode-and-Forward (DF), and Compress-and-Forward (CF) to TWRC. AF strategy \cite{acssc.rw05} is a linear relaying protocol where the
relay only scales the received signal to meet its power constraints. This simple strategy suffers from noise amplification especially at low signal-to-noise ratios (SNRs). With DF strategy, the relay jointly decodes both messages, and then re-encodes them before broadcasting the resulting codeword. 
The authors in \cite{acssc.rw05} derived an achievable rate region for TWRC by using DF strategy and superposition coding in the BC phase. This region has been improved in \cite{zsc.k06} where the authors propose that the relay sends a modulo sum of the decoded messages, thus 
mimicking the initial example of XOR NC. These DF relaying based schemes require full decoding of the incoming signals and thus suffer from a multiplexing loss due to the MAC phase limitation \cite{gdrisis.k07}.

The authors in \cite{mobicom.zll06, ieeeIT.wnps10} propose PNC schemes based on a partial DF (pDF) where the relay does not decode completely the incoming signals, but relies on the side information available at each terminal to decode a linear function of the transmitted codewords. The key strategy in these schemes is to design the codes at both transmitting terminals in the MAC phase so that the relay can compute a message which is decodable by both nodes during the BC phase. 
Nested lattice codes, which have the nice property to ensure that any integer-valued linear combination of codewords is a codeword,
are used in \cite{ieeeIT.wnps10} to implement pDF for Gaussian channels.
However, the problem of pDF schemes is to guarantee phase coherence at the relay during the MAC channel \cite{gdrisis.k07}.

Another strategy is based on the relay compressing its observation and sending it to the sources, utilizing Wyner-Ziv binning.
This strategy has attracted particular attention since it offers a good trade-off between processing complexity at the relay and noise 
amplification. CF for TWRC \cite{isit.rw06} follows the same approach as CF schemes for the relay channel \cite{ieeeIT.cg79}.
Performance bounds of CF scheme for TWRC have been investigated in \cite{acssc.sos07, sarnoff.kdmt08, allerton.gtn08}. It has been shown that for specific channel conditions, namely symmetric channels, CF outperforms the other relaying schemes at high SNR regimes. Random coding tools have been used in the aforementioned references to derive achievable rate regions of CF. Structured codes, on the other hand, have been found to be more advantageous in practical settings thanks to their reduced implementation complexity \cite{book.conway98}.

In \cite{atc.skszd12}, we have proposed a CF scheme that is based on nested lattice coding. In the MAC phase of this scheme, the communicating nodes simultaneously send their messages and the relay receives a mixture of the transmitted signals. 
The relay considers this mixture as a source which is compressed
and transmitted during the BC phase. Taking into account that each terminal has
a partial knowledge of this source (namely, its own signal that has been transmitted during the MAC phase, now considered as receiver side information),
the BC phase is
equivalent to a Wyner-Ziv compression setting with two decoders, each one having its own side information.
Each user employs lattice decoding technique to retrieve its data based on the available side information.
The proposed scheme can be seen as an extension of lattice quantization introduced in \cite{ieeeIT.zse02} to the TWRC model.
In this paper, we first generalize this latter scheme and we apply the results to our transmission problem.

In the simplest situation, when a single "layer" of compression is performed, the relay broadcasts a common compressed message to both terminals.
Therefore it is easily understood that the achievable rates in both directions are somewhat constrained by the capacity of the worst channel.
In this case, the user experiencing better channel and side information conditions 
is strongly constrained by this restriction on its transmission rate. To overcome this limitation,
in an improved scheme, the relay also sends an individual description of its output that serves as an enhancement compression layer to be recovered only by the best receiver.
Therefore, the new scheme employs three 
nested lattices. The common information is encoded using two nested lattices while the refinement information is encoded with a finer lattice that contains the other two lattices. The channel codewords corresponding to the two layers are superimposed and sent during the BC phase. 
Through numerical analysis, we show that this layered scheme outperforms AF and CF strategies in all SNR regimes and DF strategy for specific SNR regions.

Layered coding for Wyner-Ziv problem has been addressed in \cite{ieeeIT.ntg10} for lossy transmission over broadcast channel with degraded side information.
In \cite{allerton.gtn08}, the authors derive the achievable rate region of layered CF coding for TWRC, based on a random coding approach.
The authors in \cite{ieeeIT.ncl10} and \cite{ieeeWC.twym12} proposed schemes for TWRC based on doubly nested lattice coding where different power constraints at all nodes are assumed. In these schemes, each of the two end terminals employs a different code (with carefully chosen rate) constructed from the lattice partition chain. The relay decodes a modulo-lattice sum of the
transmitted codewords from the received signal. However, in \cite{ieeeIT.ncl10} full-duplex nodes are considered and in \cite{ieeeWC.twym12}, the direct link between both terminals is exploited and the transmission is performed in three phases. In these schemes,  
the relay follows a pDF strategy since it decodes a function of the transmitted lattice codewords. On the other hand, in our proposed enhancement scheme, doubly nested lattice coding is only employed at the relay for CF strategy and half-duplex terminals are considered with no direct link between the two end terminals. Furthermore, the relay does not need to know neither the other terminals' codebooks nor the precise value of the channel. It merely
reconstructs its encoder from the channel module and the variances of the transmitted signals. To our knowledge, our work is the first that proposes a doubly nested lattice coding for CF relaying in TWRC.

The remaining of the paper is organized as follows. Section \ref{Sec::sysmodel} introduces the system model. Section \ref{Sec::onelayer} derives the achievable rate region when one layer lattice-based coding scheme is used and section \ref{Sec::twolayers} derives the achievable rate region with two layer lattice-based coding. Section \ref{Sec::sims} illustrates the performance of the proposed schemes through numerical results. Finally, section \ref{Sec::concl} concludes the paper.

\noindent \textbf{Notations:} 
Random variables (r.v.) are indicated by capital letters and their realizations are denoted by small letters. Vector of r.v. or a sequence of realizations are indicated by bold fonts.
\section{System Model}
\label{Sec::sysmodel}
\begin{figure}[htbp]
\centering
\includegraphics[width=0.5\linewidth]{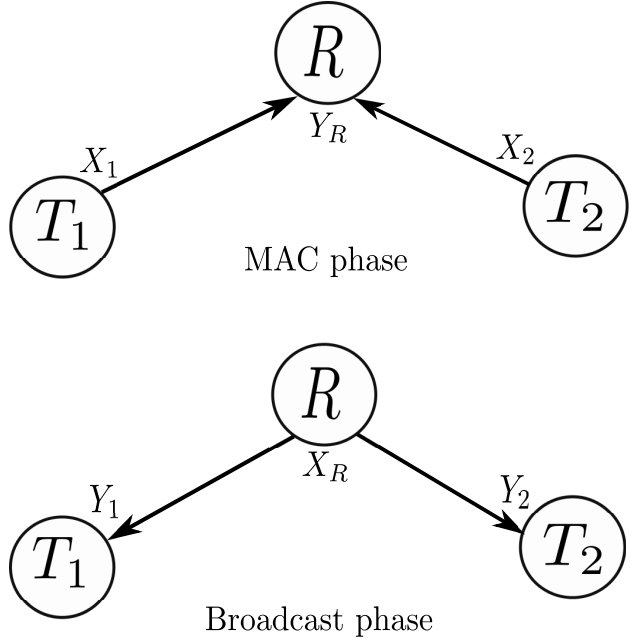}
\caption{The two-phase transmission of TWRC: MAC and Broadcast phases}\label{Fig1}
\end{figure}
Consider a Gaussian TWRC in which two source nodes T$_1$ and
T$_2$ exchange two individual messages $m_1$ and $m_2$, with the help of a
relay $R$ as shown in Fig.\ref{Fig1}. For this model, we have
the following assumptions: 
\begin{enumerate}[label={a.\arabic*}]
\item There is no direct link between T$_1$ and T$_2$.
\item The relay and the source nodes operate in half-duplex mode;
\item The communication takes $n$ channel uses that are split into two orthogonal phases: MAC phase and BC phase with lengths $n_1 = \alpha n$ and $n_2 = (1- \alpha) n $ , $\alpha \in [0, 1]$ respectively.
\end{enumerate}
During the MAC phase, node T$_1$ draws uniformly a message $m_1$ from the set $\mathcal{M}_1 =\{1, 2,\cdots ,2^{n R_{12}}\}$ and sends it to the other terminal T$_2$ where $R_{12}$ denotes the message rate of node T$_1$ destined to T$_{2}$. Similarly, node T$_2$ draws uniformly a message $m_2$ from the set $\mathcal{M}_2 =\{1, 2,
\cdots ,2^{n R_{21}} \}$ and sends it to the other terminal T$_1$ where $R_{21}$ denotes the message rate of node T$_2$ destined to T$_{1}$. Let $\mathbf{x}_i(m_i) \in \mathbb{R}^{n_1}$ be the channel codeword of length $n_1$ sent by node T$_i$, $i = 1, 2$ and $P_i$ be the corresponding transmit power constraint that verify the following assumptions
\begin{enumerate}[resume,label={a.\arabic*}]
\item $\frac{1}{n_1} E[||\mathbf{X}_i||^2] \leq P_i $
\end{enumerate}
The messages are transmitted through a memoryless Gaussian channel 
and the relay R receives a signal $\mathbf{y}_R \in \mathbb{R}^{n_1}$ given by 
\begin{equation}
\mathbf{y}_R =  h_1 \mathbf{x}_1 + h_2 \mathbf{x}_2 + \mathbf{z}_R
\label{Yr}
\end{equation}
where $h_{i}$ denotes the channel coefficient between T$_i$ and R, $i = 1, 2$. We assume that:
\begin{enumerate}[resume,label={a.\arabic*}]
\item The components of the random vector $\mathbf{Z}_R$ are i.i.d Additive White Gaussian Noise (AWGN) at the relay with variance $\sigma_R^2$ i.e. $\sim \mathcal{N}(0,\sigma_R^2)$ and they are independent from the channel inputs $\mathbf{X}_i$, $i = 1, 2$.
\item The channel coefficients follow a block fading model. 
Without loss of generality, channel reciprocity between MAC and BC channels is assumed, i.e. $h_{i \rightarrow R} = h_{R \rightarrow i} = h_i$.
\end{enumerate}
During the BC phase, the relay generates a codeword $\mathbf{x}_R(m_R) \in \mathbb{R}^{n_2}$ of dimension $n_2$ from the received sequence $\mathbf{y}_R$. The average power constraint at the relay $P_R$ verifies
\begin{enumerate}[resume,label={a.\arabic*}]
\item $\frac{1}{n_2} E[||\mathbf{X}_R||^2] \leq P_R $
\end{enumerate}
The signal $\mathbf{x}_R$ is transmitted through a broadcast memoryless channel
and the received signal at node T$_i$ is $\mathbf{y}_i \in \mathbb{R}^{n_2}$, $i = 1, 2$.
\begin{equation}
\mathbf{y}_i = h_i \mathbf{x}_R + \mathbf{z}_i \label{Y1},
\end{equation}
\begin{enumerate}[resume,label={a.\arabic*}]
\item The components of $\mathbf{Z}_i$ are i.i.d AWGN at node T$_i$ with variance $\sigma_i^2$, $i= 1, 2$ and they are independent from the channel input $\mathbf{X}_R$.
\end{enumerate}
Perfect CSI is assumed at all nodes. This assumption will be discussed more in detail in \textsl{Remark \ref{remarkFin}}.
\noindent For the aforementioned TWRC, a rate pair $(R_{12}, R_{21})$ is said to be achievable if there exists a sequence of encoding and decoding functions such that the decoding error probability approaches zero for $n$ sufficiently large.

For the sake of completeness, we hereafter outline some preliminaries on lattices \cite{book.conway98, ieeeIT.ez04}.
\subsection*{Fundamentals on Lattice Coding:}
\label{SubSec::lcodes}
A real $n_1$-dimensional lattice $\Lambda$ is a subgroup of the Euclidean space $(\mathbb{R}^{n_1},+)$. $\forall \lambda_1, \lambda_2 \in \Lambda$, $\lambda_1+\lambda_2 \in \Lambda$. We present below some fundamental properties associated with a lattice:
\begin{itemize}
\item The nearest neighbor lattice quantizer of $\Lambda$ is defined as $Q_{\Lambda}(\mathbf{x}) = \displaystyle \arg\min_{\lambda\in \Lambda} ||\mathbf{x}- \lambda||$ where $\textbf{x} \in \mathbb{R}^{n_1}$ and $\|.\|$ is the Euclidean norm.
\item The basic Voronoi cell of $\Lambda$ is the set of points in $\mathbb{R}^{n_1}$ closer to the zero vector than to any other point of $\Lambda$ , $\mathcal{V}(\Lambda) = \{\textbf{x} ~|~ Q_\Lambda(\textbf{x}) = \textbf{0}\}$.
\item The volume of a lattice $V := \text{Vol}(\mathcal{V}(\Lambda))$.
\item The mod-$\Lambda$ operation is defined as \textbf{x} mod $\Lambda = \textbf{x} - Q_{\Lambda}(\textbf{x})$. It satisfies the distributive law: (\textbf{x} mod $\Lambda + \textbf{y})\mod \Lambda = (\textbf{x} + \textbf{y}) \mod \Lambda$.
\item The second moment per dimension of $\Lambda$ is $\sigma^2(\Lambda) :=  \frac{1}{n_1}.\frac{1}{V} \int_{\mathcal{V}(\Lambda)} ||\textbf{x}||^2 d\textbf{x}$.
\item The dimensionless normalized second moment is defined as $G(\Lambda):= \frac{\sigma^2(\Lambda)}{V^{2/{n_1}}} $.
\item A sequence of $n_1$-dimensional lattices $\Lambda^{(n_1)}$ is said to be good for quantization if $G(\Lambda^{(n_1)}) \underset{n_1 \rightarrow \infty}{\longrightarrow} \frac{1}{2 \pi e}$ \cite{ieeeIT.zf96}.
\item A sequence of $n_1$-dimensional lattices $\Lambda^{(n_1)}$ is said to be good for AWGN channel coding if for $n_1$-dimensional vector $\mathbf{Z} \sim \mathcal{N}(\mathbf{0},\sigma^2 \mathbf{I}_{n_1})$, $P\{\mathbf{Z} \notin \mathcal{V}(\Lambda^{(n_1)})\}$ vanishes when $n_1$ goes to $\infty$. In this case, $\text{Vol}(\Lambda^{(n_1)}) \underset{n_1 \rightarrow \infty}{\longrightarrow} 2^{n_1 h(Z)}$, where $h(\mathbf{Z}) = \frac{1}{2} \log(2 \pi e \sigma^2)$ is the differential entropy of $\mathbf{Z}$ \cite{ieeeIT.poltyrev94}. 
\item There exist lattices which are simultaneously good for quantization and channel coding in \cite{ieeeIT.elz05}.
\item \begin{lemma} \label{CryptoL} Crypto Lemma \cite{ieeeIT.ez04}. For a dither vector $\textbf{T}$ independent of $\textbf{X}$ and uniformly distributed over $\mathcal{V}(\Lambda)$, then $\textbf{Y}= (\textbf{X} + \textbf{T})\mod \Lambda$ is uniformly distributed over $\mathcal{V}(\Lambda)$ and is independent of $\textbf{X}$.\end{lemma} 
\end{itemize}

\noindent Consider a pair of $n_1$-dimensional nested lattices $(\Lambda_1, \Lambda_2)$ such as $\Lambda_2 \subset \Lambda_1$. The fine lattice is $\Lambda_1$ with basic Voronoi region $\mathcal{V}_1$ of volume $V_1$ and second moment per dimension $\sigma^2(\Lambda_1)$. The coarse lattice is $\Lambda_2$ with basic Voronoi region $\mathcal{V}_2$ of volume $V_2$ and second moment $\sigma^2(\Lambda_2)$. The following properties of nested lattices hold:
\begin{itemize}
\item For $\Lambda_2 \subset \Lambda_1$, we have $Q_{\Lambda_2}(Q_{\Lambda_1}(x)) = Q_{\Lambda_1}(Q_{\Lambda_2}(x)) = Q_{\Lambda_2}(x)$.
\item The points of the set ${\Lambda_1 \cap \mathcal{V}_2} = \Lambda_1\mod \Lambda_2$ represent the coset leaders of $\Lambda_2$ relative to $\Lambda_1$, where for each $\lambda \in \{\Lambda_1 \mod \Lambda_2\}$, the shifted lattice $\Lambda_{2,\lambda} =  \Lambda_{2} +\lambda$ is called a coset of $\Lambda_2$ relative to $\Lambda_1$. There are $\displaystyle \frac{V_2}{V_1}$ distinct cosets. It follows that the coding rate when using nested lattices is 
\begin{equation}
\label{codR}
R = \frac{1}{n_1} \log_2 |{\Lambda_1 \cap \mathcal{V}_2}| = \frac{1}{n_1} \log_2 \frac{V_2}{V_1} ~~\text{(bits per dimension)}.
\end{equation}
\end{itemize}
\section{Achievable Rate Region for TWRC}
\label{Sec::onelayer}
\begin{theorem} \label{Theorem1}
For a Gaussian TWRC, under the assumptions a.1 to a.8, the convex hull of the following end-to-end rate-pairs $(R_{12}, R_{21})$ is achievable:
\begin{eqnarray}
\label{R12}
R_{12} & \leq \displaystyle \frac{\alpha}{2} \log_2\left( 1 + \displaystyle \frac{|h_1|^2 P_1}{\sigma_R^2  + \displaystyle \frac{|h_1|^2 P_1 + \sigma_R^2}{\left(1+\displaystyle \min_{i \in \{1,2\}}\frac{|h_i|^2 P_R}{\sigma_i^2}\right)^{\frac{1 - \alpha}{\alpha}} -1 }} \right) \\ 
R_{21} & \leq \displaystyle \frac{\alpha}{2} \log_2\left( 1 + \displaystyle \frac{|h_2|^2 P_2}{\sigma_R^2  + \displaystyle  \frac{|h_1|^2 P_1 + \sigma_R^2}{\left(1+ \displaystyle \min_{i \in \{1,2\}}\frac{|h_i|^2 P_R}{\sigma_i^2}\right)^{\frac{1 - \alpha}{\alpha}} - 1}} \right) \label{R21}
\end{eqnarray}
for $\alpha \in [0,1]$.
\end{theorem}
\vspace*{10mm}

The main idea of the proposed scheme is the following: during the BC phase, the relay sends a quantized version of the signal that was received during the MAC phase. It uses nested lattices to generate a source index that is then channel encoded. This index is decoded by both users and, based on their own information (sent during the MAC phase), the sources recover each the message which is sent to them.  
The proof of \textsl{Theorem \ref{Theorem1}} is detailed in the next paragraphs: in section \ref{SubSec:coding}, the lattice coding scheme for the source coding is presented. The end-to-end achievable rates are derived in section \ref{Subsec::rateAnalysis} and finally in section \ref{Subsec::achievblRates} 
the achievable rate region is maximized by appropriate optimization of lattice parameters.
\subsection{Lattice Based Source Coding}
\label{SubSec:coding}
We suppose that the elements of $\mathbf{X}_i$, $i = 1, 2$, are drawn from an independent identically distributed (i.i.d) Gaussian distribution with zero mean and variance $P_i$.
Let $\mathbf{S}_i = h_i \textbf{X}_i$ be the side information available at terminal T$_i$, $i = 1, 2$. 
The signal sent by the relay $\mathbf{Y}_R$ can be written in two ways as the sum of two independent Gaussian
r.v.: the side information $\mathbf{S}_i$ and the unknown part
$\mathbf{U}_i = \mathbf{Y}_R |\mathbf{S}_i = h_{\bar{i}} \mathbf{X}_{\bar{i}} +
\mathbf{Z}_R$, $i \in \{1,2\}$. From their received signals, each terminal T$_i$ , $i \in \{1,2\}$ decodes $\mathbf{\hat{U}}_i$ using $\mathbf{S}_i$. The variance per dimension of $\textbf{U}_i$ is $\sigma_{U_i}^2 = VAR(Y_R |S_i) = |h_{\bar{i}}|^2 P_{\bar{i}} +\sigma_R^2$.\\
In the following, we detail the proposed lattice source coding scheme. 
\subsubsection{Encoding}
\begin{figure}[htbp]
\centering
\includegraphics[width =\linewidth]{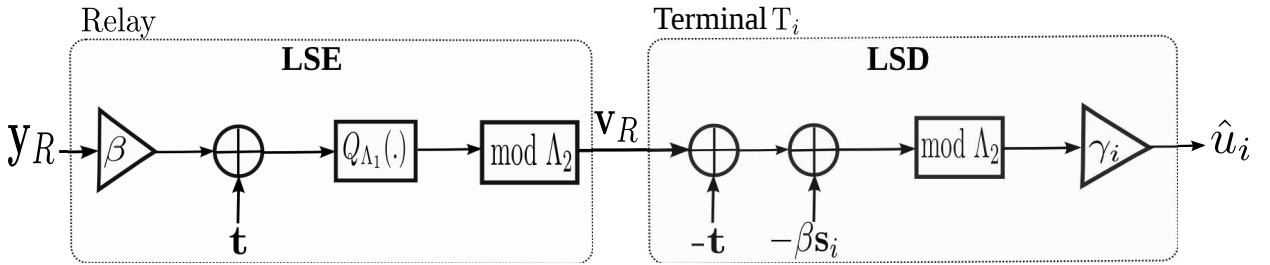}
\caption{Lattice encoding at the relay and decoding at T$_i$, $i = 1, 2$ }\label{Fig2}
\end{figure}

The lattice source encoding (LSE) operation is performed with four successive operations:
first, the input signal ${\bf y}_R$ is scaled with a factor $\beta$. Then, a
random dither $\textbf{t}$ which is uniformly distributed over $\mathcal{V}_{1}$ is
added. This dither is known by all nodes. The dithered scaled version of ${\bf y}_R$, $\beta \textbf{y}_R + \textbf{t}$ is quantized to the nearest
point in $\Lambda_1$. The outcome of this operation is processed
with a modulo-lattice operation in order to generate a vector
$\textbf{v}_R$ of size $n_1$ as shown in Fig.\ref{Fig2}, and defined by:
\begin{equation}
\label{vr}
\textbf{v}_R = Q_{\Lambda_1}(\beta \textbf{y}_R + \textbf{t})\mod\Lambda_2.
\end{equation}
The relay sends the index of $\textbf{v}_R$ that identifies a coset of $\Lambda_2$ relative to $\Lambda_1$ that contains $Q_{\Lambda_1}(\beta \textbf{y}_R + \textbf{t})$. By construction, the coset leader $\textbf{v}_R$ can be represented using $\displaystyle \log_2\left(\frac{V_2}{V_1}\right)$
bits. Thus, the rate of the source encoding scheme employed by the relay is $R$ given by Eq. \eqref{codR}. We assume further that $\Lambda_1$ is good for quantization and $\Lambda_2$ is good for channel coding \cite{ieeeIT.zse02}. For high dimension $n_1$ and according to the properties of good lattices, we have $\frac{1}{n_1}\log_2(V_i) \approx \frac{1}{2} \log_2(2 \pi e \sigma^2(\Lambda_i))$ , $i \in \{1,2\}$. Thus $R$ reads  
\begin{equation}
\label{FRateD} 
R = \displaystyle \frac{1}{2} \log_2\left(\frac{\sigma^2(\Lambda_2)}{\sigma^2(\Lambda_1)}\right). 
\end{equation}
\subsubsection{Decoding}
For both users, $\textbf{v}_R$ is decoded first. Then $\hat{\textbf{u}}_i$ is reconstructed with a lattice source decoder (LSD) using the side information $\textbf{s}_i$ as 
\begin{equation}
\label{ui}
\hat{\textbf{u}}_i = \gamma_i((\textbf{v}_R -\textbf{t} - \beta \textbf{s}_i) \mod\Lambda_2), ~ i = 1, 2
\end{equation} 
where $\gamma_i$, $i \in \{1,2 \}$ are the scaling factors at each decoder.
\subsection{Rate Analysis}
\label{Subsec::rateAnalysis}
At the relay, message $m_R$ corresponding to the index of
$\textbf{v}_R$ is mapped to a codeword $\textbf{x}_R$ of size $n_2$. We assume that the elements of the r.v. $\mathbf{X}_R$ are drawn from an i.i.d Gaussian distribution with zero mean and variance $P_R$. The broadcast rate from the relay to both terminals is bounded by the capacity of the worst individual relay-terminal channel capacity $\min(I(X_R; Y_1), I(X_R; Y_2))$. From  Shannon's source-channel separation theorem \cite{book.cover91}, we have 
\begin{equation}
\label{SCST}
n_1 R\leq n_2 \min(I(X_R; Y_1), I(X_R; Y_2)).
\end{equation}
Since real Gaussian codebooks are used for all transmissions, 
we have: $I(X_R; Y_i)= \frac{1}{2}\log_2\left(1 + \frac{|h_i|^2 P_R}{\sigma_i^2}\right)$, $i= 1,2$.
Finally, by combining Eq. \eqref{FRateD} and \eqref{SCST}, we obtain the following constraint on the achievable rates
\begin{align}
\label{CondR}
n_1 \displaystyle \log_2\left(\frac{\sigma^2(\Lambda_2)}{\sigma^2(\Lambda_1)}\right) \leq  n_2 \log_2 \left(1+ \displaystyle \min_{i \in \{1,2\}} \frac{|h_i|^2 P_R}{\sigma_i^2}\right).
\end{align}
This constraint ensures that index $m_R$ is transmitted reliably to both terminals and $\textbf{v}_R$ is available at the input of the LSD of both receivers. 
At terminal T$_i$, $\hat{\textbf{u}}_{i}$ in \eqref{ui} can be written as:
\begin{eqnarray}
\label{u2}
&\hat{\textbf{u}}_i & =   \gamma_i ((\beta \textbf{u}_i +\textbf{e}_q )~\text{mod}~ \Lambda_2 )\label{lgn3} \\
& 				   & =   \gamma_i (\beta \textbf{u}_i + \textbf{e}_q )\label{lgn4}						
\end{eqnarray} 
where $\textbf{e}_q = Q_{\Lambda_1}(\beta \textbf{y}_R + \textbf{t}) - (\beta \textbf{y}_R + \textbf{t}) = -(\beta \textbf{y}_R + \textbf{t}) \mod\Lambda_1$, is the quantization error. By \textsl{Lemma \ref{CryptoL}}, $\textbf{E}_q$ is independent from $\textbf{Y}_R$, and thus from $\textbf{U}_i$. Also $\textbf{E}_q$ is uniformly distributed over $\mathcal{V}_1$ thus the variance of $\textbf{E}_q$ per dimension is $\sigma^2(\Lambda_1)$.
Equation \eqref{lgn4} is valid only if
$\beta \textbf{u}_i +\textbf{e}_q \in \mathcal{V}_2$. According to \cite{ieeeIT.zse02}, with good channel coding lattices, the probability $\text{Pr}(\beta
\mathbf{U}_i  + \mathbf{E}_q \notin \mathcal{V}_{2})$ vanishes asymptotically provided that:
\begin{equation}
\frac{1}{n_1} \mathbb{E}\|\beta \mathbf{U}_i  + \mathbf{E}_q \|^2 = \beta^2 \sigma^2_{U_i} + \sigma^2(\Lambda_1) \leq \sigma^2(\Lambda_2)
\label{consS}
\end{equation}
By replacing $\textbf{U}_i$ by its value, we conclude that:
\begin{equation}
\hat{\textbf{U}}_i 
 = \gamma_i( \beta( h_1 \textbf{X}_1 + \textbf{Z}_R) + \textbf{E}_q ).
 \label{virtualchani}
\end{equation}
Let $\textbf{Z}_{eq,i} = \gamma_i (\beta \textbf{Z}_R + \textbf{E}_q)$ be the
effective additive noise at terminal T$_i$. For high dimension assumption, $n_1 \rightarrow \infty$, we can
approximate the uniform variable $\mathbf{E}_q$ over $\mathcal{V}_1$ by a Gaussian variable $\mathbf{Z}_q$ with the same variance \cite{ieeeIT.zf96}. Therefore, the communication between terminals T$_1$ and T$_2$ (resp. T$_2$ and T$_1$) is equivalent to a AWGN channel where the Gaussian noise is given by $\textbf{Z}_{eq,i}$.
hence, the achievable rates of both links satisfy
\begin{eqnarray}
\label{R12tmp}
n R_{12} &\leq \displaystyle \frac{n_1}{2} \log_2\left(1 + \displaystyle\frac{\beta^2 |h_1|^2 P_1}{\beta^2 \sigma_R^2 + \sigma^2(\Lambda_1)}\right)  \\
n R_{21} &\leq \displaystyle\frac{n_1}{2} \log_2\left(1 + \displaystyle\frac{\beta^2 |h_2|^2 P_2}{\beta^2 \sigma_R^2 + \sigma^2(\Lambda_1)}\right) \label{R21tmp}
\end{eqnarray}
\subsection{Achievable Rate Region}
\label{Subsec::achievblRates}
The rate region that can be achieved by the proposed scheme is characterized by the constraints \eqref{R12tmp}, \eqref{R21tmp}, \eqref{CondR} and \eqref{consS}. Without loss of generality, we assume that $|h_2|^2 P_2 \leq |h_1|^2
P_1$. With this setting, T$_2$ is the terminal which experiences the weakest side information. Letting $\alpha = \displaystyle \frac{n_1}{n}$, from \eqref{CondR} and \eqref{consS}, the lower bound of $\sigma^2(\Lambda_1)$ is given by
\begin{equation}
\label{LBsigma1}
\sigma^2(\Lambda_1) \geq \frac{\beta^2 \sigma^2_{U_2}}{\left(1+ \displaystyle \min_{i \in \{1,2\}} \frac{|h_i|^2 P_R}{\sigma_i^2}\right)^{\frac{1 -\alpha}{\alpha}} -1}
\end{equation}

The rate region defined in \eqref{R12tmp} and \eqref{R21tmp} can be rewritten as
\begin{eqnarray}
R_{12} \leq \frac{\alpha}{2} \log_2\left(1 + \mbox{SNR}_{1 \rightarrow 2} \right)\\
R_{21} \leq  \frac{\alpha}{2} \log_2\left(1 +\mbox{SNR}_{2 \rightarrow 1} \right)
\end{eqnarray}
where SNR$_{1 \rightarrow 2}$ and SNR$_{2 \rightarrow 1}$ are the end-to-end SNRs, defined as follows:
\begin{eqnarray}
\label{SNR12}
\mbox{SNR}_{1 \rightarrow 2} = \frac{\beta^2 |h_1|^2 P_1}{\beta^2 \sigma_R^2 + \sigma^2(\Lambda_1)} \\
\mbox{SNR}_{2 \rightarrow 1} = \frac{\beta^2 |h_2|^2 P_2}{\beta^2 \sigma_R^2 + \sigma^2(\Lambda_1)} \label{SNR21}
\end{eqnarray}
We notice that SNR$_{1 \rightarrow 2}$ and SNR$_{2 \rightarrow 1}$ are maximized when $\sigma^2(\Lambda_1)$ is minimal. Thus 
the optimal choice on the second moment of $\Lambda_1$ is 
\begin{equation}
\label{sigmaMim}
\sigma^2(\Lambda_1)_{\min} = \frac{\beta^2 \sigma^2_{U_2}}{\left(1+ \displaystyle \min_{i \in \{1,2\}} \frac{|h_i|^2 P_R}{\sigma_i^2}\right)^{\frac{1 -\alpha}{\alpha}} -1}
\end{equation}
Finally by replacing  $\sigma^2(\Lambda_1)_{\min}$ in \eqref{SNR12} and \eqref{SNR21}, Eq. \eqref{R12} and \eqref{R21} are verified and the proof is concluded.
\begin{remark}
For the transmission problem of the TWRC, the achievable rate region is independent of the choice of the decoders scaling factors $\gamma_i$. It is also independent of the encoder scaling factor $\beta$ provided that $\sigma^2(\Lambda_1)$ is set to its smallest value $\sigma^2(\Lambda_1)_{\min}$ in \eqref{sigmaMim}. 
In the next section, we show that these parameters that are involved in the source coding problem that was addressed in \cite{atc.skszd12}.
\end{remark}
\subsection{Analog Signal Transmission}
\label{Subsec::distr}
When using the relay to transmit analog signals, the distortion that affects the reconstructed signals becomes the main performance metric. The second moment of this distortion is given by
\begin{equation}
\label{Distv1}
\frac{1}{n_1} \mathbb{E}\|\mathbf{Y}_R - \mathbf{\hat{Y}}_{Ri}\|^2 = D_i ~;~ i \in \{1,2\}
\end{equation}
where $\mathbf{Y}_R = \mathbf{U}_i + \mathbf{S}_i$ and $\mathbf{\hat{Y}}_{Ri} = \mathbf{\hat{U}}_i + \mathbf{S}_i$. By replacing $\mathbf{\hat{U}}_i$ by its value in \eqref{lgn4}, \eqref{Distv1} becomes
\begin{equation}
\label{Distv2}
D_i = (1 - \gamma_i \beta)^2 \sigma^2_{U_i} + \gamma^2_i \sigma^2(\Lambda_1) ~;~ i \in \{1,2\}.
\end{equation}
For the analog signal transmission, this distortion has to be minimized to obtain the optimal source coding scheme. For fixed $\beta$, the distortion at T$_i$ depends only on two parameters namely $\gamma_i$ and $\sigma^2(\Lambda_1)$.
The optimal distortion can be obtained by calculating the following derivatives:
\begin{subequations}
\begin{align}
\label{dervts}
\frac{\partial D_i}{\partial \gamma_i} = 0 &\Rightarrow  \gamma_i^{*} = \frac{\beta \sigma^2_{U_i} }{\beta^2 \sigma^2_{U_i} + \sigma^2_{\Lambda_1} } \\
\frac{\partial D_i}{\partial \sigma^2(\Lambda_1)} = 0 &\Rightarrow \gamma_i^{*} = 0
\end{align}
\end{subequations}
where $\gamma_i^{*}$, $i \in \{1,2 \}$ are the optimal decoder scaling factors. Since $\gamma_i > 0$, then $\frac{\partial D_i}{\partial \sigma^2(\Lambda_1)} > 0$. Thus, the function $D_i$ is increasing with $\sigma^2(\Lambda_1)$ and $\sigma^2(\Lambda_1)_{\min}$ in \eqref{sigmaMim} is the optimal choice that minimizes the distortion at each terminal. Therefore, 
\begin{equation}
\label{gammaistar}
\gamma_i^{*} = \frac{\beta \sigma^2_{U_i} }{\beta^2 \sigma^2_{U_i} + \sigma^2(\Lambda_1)_{\min}} ~, ~ i \in \{1,2 \}.
\end{equation}
By replacing $\sigma^2(\Lambda_1)$ and $\gamma_i$ by their optimal values, we obtain the minimal value of $D_i^{\min}$ given by
\begin{align}
D_i^{\min} & = \frac{\sigma^2(\Lambda_1)_{\min} \sigma^2_{U_i} }{\beta^2 \sigma^2_{U_i} + \sigma^2(\Lambda_1)_{\min}} \label{Dimin} \\
 & = \frac{\sigma^2_{U_2} \sigma^2_{U_i} }{\left(\left(1+ \displaystyle \min_{i \in \{1,2\}} \frac{|h_i|^2 P_R}{\sigma_i^2}\right)^{\frac{1 -\alpha}{\alpha}} - 1 \right)\sigma^2_{U_i} +  \sigma^2_{U_2}} ~,~ i \in \{1,2 \}.
\end{align}
$D_i^{\min}$, $i \in \{1,2 \}$, just like the achievable rates, are independent of $\beta$. However, for a fixed $\beta$, the lattice parameters and receivers scaling factors depend on that choice.
\subsubsection*{Comments on the Distortions}
At terminal T$_2$, 
the distortion writes:
\begin{eqnarray*}
&D_2^{\min} & = \frac{\sigma^2_{U_2} \sigma^2_{U_2} }{(A- 1) \sigma^2_{U_2} + \sigma^2_{U_2}} \\
&  & =  \frac{\sigma^2_{U_2}}{A}
\end{eqnarray*}
where $A = \left(1+ \displaystyle \min_{i \in \{1,2\}} \frac{|h_i|^2 P_R}{\sigma_i^2}\right)^{\frac{1 -\alpha}{\alpha}}$. It can be reformulated as
\begin{equation*}
\frac{\sigma^2_{U_2}}{D_2^{\min}} =  \left(1+ \displaystyle \min_{i \in \{1,2\}} \frac{|h_i|^2 P_R}{\sigma_i^2}\right)^{\frac{1 -\alpha}{\alpha}}
\end{equation*}
\begin{equation}
\label{ConsApp}
\alpha \log_2\left(\frac{\sigma^2_{U_2}}{D_2^{\min}}\right) = (1 - \alpha) \log_2\left(1+ \displaystyle \min_{i \in \{1,2\}} \frac{|h_i|^2 P_R}{\sigma_i^2}\right)
\end{equation}
We find, in the left hand side of Eq. \eqref{ConsApp}, the Wyner-Ziv rate distortion function of the Gaussian source $\textbf{Y}_R$ with side information $\textbf{S}_2$ at the decoder T$_2$ \cite{if.wz78}. 
It is defined as the minimum rate needed to achieve $D_2^{\min}$ and it is given by:\begin{equation} 
R_{WZ}(D_2^{\min}) = \frac{1}{2} \log_2\left(\frac{\sigma^2_{U_2}}{D_2^{\min}}\right)
\end{equation}
Note that the source coding rate is no larger than the channel coding rate to the relay.
Also, according to \eqref{gammaistar} the optimal value of $\gamma_2$ is given by
\begin{eqnarray*}
\label{gamma2star}
& \gamma_2^{*} & = \frac{\beta \sigma^2_{U_2}}{\beta^2 \sigma^2_{U_2} + \sigma^2(\Lambda_1)_{\min}} \\
\end{eqnarray*}
With the choice $\beta = \gamma_2^{*}$, we get $\beta = \sqrt{1  - \frac{D_2^{\min}}{\sigma^2_{U_2}}}$. 
This is in accordance with the optimal scaling factor reported in \cite{if.wz78, atc.skszd12} for the optimum Gaussian forward test channel. For this choice of $\beta$,
$\sigma^2(\Lambda_1)_{\min} = D_2^{\min}$ which is consistent with the source coding parameters choices in \cite{atc.skszd12}.

At terminal T$_1$, the reconstruction distortion is smaller than $D_2^{\min}$ of terminal T$_2$. This is compatible with the fact that T$_1$ has the best side information quality and the proposed achievable scheme is optimal for the worst user.
\section{Improved Achievable Rate Region for TWRC}
\label{Sec::twolayers}
In the previous section, we presented a PNC scheme in which a common information is sent from the relay to both users. The rates that are achievable by this scheme depend only on the ratio $\frac{\sigma^2(\Lambda_1)_{\min}}{\beta^2}$. 
This ratio is determined, as shown by Eq.\eqref{sigmaMim}, essentially by 
the variance $\sigma^2_{U_i}$ of the unknown part of the source at the terminal T$_i$ and the lowest channel coefficient amplitude $ \displaystyle \min_{i \in \{1,2\}} \frac{|h_i|^2}{\sigma_i^2} $. Thus, the achievable rates are limited by the user which has the weakest side information and also the worst channel condition.
In this case, the best user suffers from this limitation on its achievable rate. In order to improve its rate, an additional refinement information can be sent from the relay, that can be only decoded by the best user.\\
Without loss of generality, let terminal T1 has a better channel condition than T2, and also more transmit power i.e $|h_1| \geq |h_2|$ and $P_1 \geq P_2$.
The following theorem provides an achievable rate region for the TWRC, obtained using the refinement scheme.

\begin{theorem}
\label{Theorem2}
For a Gaussian TWRC, under the assumptions a.1 to a.8, the convex hull of the following end-to-end rate-pairs $(R_{12}, R_{21})$ is achievable:
\begin{align}
\label{R12ref}
R_{12} & \leq \displaystyle \frac{\alpha}{2} \log_2\left( 1 + \displaystyle \frac{|h_1|^2 P_1}{\sigma_R^2  + \displaystyle \frac{|h_1|^2 P_1 + \sigma_R^2}{\left(1+ \displaystyle \frac{\nu |h_2|^2 P_R}{(1 -\nu) |h_2|^2 P_R + \sigma_2^2}\right)^{\frac{1 - \alpha}{\alpha}} -1}} \right) \\ 
R_{21} & \leq  \displaystyle \frac{\alpha}{2} \log_2\left( 1 + \displaystyle \frac{|h_2|^2 P_2}{\sigma_R^2  + \displaystyle  \frac{|h_1|^2 P_1 + \sigma_R^2}{\left(1 + \displaystyle \frac{(1 -\nu) |h_1|^2 P_R}{ \sigma_1^2}\right)^{\frac{1- \alpha}{\alpha}} \left[ \left(1+ \displaystyle \frac{\nu |h_2|^2 P_R}{(1 -\nu) |h_2|^2 P_R + \sigma_2^2}\right)^{\frac{1 - \alpha}{\alpha}} -1 \right]}} \right)
\label{R21ref}
\end{align}
for $\alpha , \nu \in [0,1]$.
\end{theorem}
\hspace*{1cm}

As we mentioned previously, the main idea of the coding scheme that we employ for Theorem 2 is having the relay sending two descriptions of its received signal, a common layer that is intended to be recovered by both users and an individual or refinement layer that is intended to be recovered by only the best user, i.e., terminal T1.\\
The proof of Theorem \ref{Theorem2} is detailed in the following subsections.
 \subsection{Doubly Nested Lattices for Source Coding}
\label{SubSec2::lcodes}
We use a doubly nested lattice chain $(\Lambda_0, \Lambda_1, \Lambda_2)$ such as $\Lambda_2 \subset \Lambda_1 \subset \Lambda_0$. We require that $\Lambda_2$ is good for channel coding, $\Lambda_1$ is simultaneously good for channel and source coding and $\Lambda_0$ is good for source coding.\\  
From these lattices, we form three codebooks
$$\mathcal{C}_c = \Lambda_1 \cap \mathcal{V}_2$$
$$\mathcal{C}_r = \Lambda_0 \cap \mathcal{V}_1 $$
$$\mathcal{C}_1 = \Lambda_0 \cap \mathcal{V}_3 $$
with the following coding rates:
\begin{eqnarray}
\label{Rcref}
R_c &= \frac{1}{n_1} \log_2\left(\frac{V_2}{V_1}\right)
\underset{n_1 \rightarrow \infty}{\longrightarrow}  \frac{1}{2} \log_2\left(\frac{\sigma^2(\Lambda_2)}{\sigma^2(\Lambda_1)}\right)\\ \label{Rrref}
R_r &= \frac{1}{n_1} \log_2\left(\frac{V_1}{V_0}\right) \underset{n_1 \rightarrow \infty}{\longrightarrow} \frac{1}{2} \log_2\left(\frac{\sigma^2(\Lambda_1)}{\sigma^2(\Lambda_0)}\right)\\
R_1 = R_c + R_r& = \frac{1}{n_1} \log_2\left(\frac{V_2}{V_0}\right) \underset{n_1 \rightarrow \infty}{\longrightarrow} \frac{1}{2} \log_2\left(\frac{\sigma^2(\Lambda_2)}{\sigma^2(\Lambda_0)}\right)  \label{R1ref}
\end{eqnarray}
where $R_c$ is the common source rate, $R_r$ is the refinement source rate and $R_1$ is the total source rate at terminal T$_1$.
\subsubsection{Encoding} 
\begin{figure}[htbp]
\centering
\includegraphics[width =0.7\linewidth]{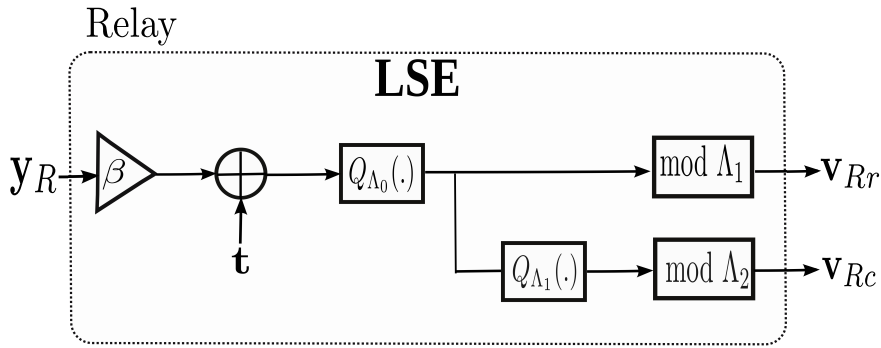}
\caption{Layered Lattice encoding at the relay}\label{Fig3}
\end{figure}

Figure \ref{Fig3} shows the LSE operation.
The input signal ${\bf y}_R$ is scaled with a factor $\beta$. Then, a random dither $\textbf{t}$ which is uniformly distributed over $\mathcal{V}_{1}$ is added. This dither is known by all nodes. The dithered scaled version of ${\bf y}_R$, $\beta \textbf{y}_R + \textbf{t}$, is quantized to the nearest
point in $\Lambda_0$. The outcome of this operation is then processed to generate two messages. First, the coset leader of $\Lambda_1$ relative to $\Lambda_0$, $\textbf{v}_{Rr}$, is generated by a modulo-lattice operation. The index of $\textbf{v}_{Rr}$ identifies the refinement message. Then, another quantization to the nearest point in $\Lambda_1$ is performed and processed with another modulo-lattice operation to generate the coset leader of $\Lambda_2$ relative to $\Lambda_1$, $\textbf{v}_{Rc}$. The index of $\textbf{v}_{Rc}$ identifies the common message.
Both messages are defines as:
\begin{equation}
\label{vrr}
\textbf{v}_{Rr} = Q_{\Lambda_0}(\beta \textbf{y}_R + \textbf{t})\mod\Lambda_1
\end{equation}
\begin{eqnarray}
&\textbf{v}_{Rc} & = Q_{\Lambda_1}( Q_{\Lambda_0}(\beta \textbf{y}_R + \textbf{t})) \mod\Lambda_2\\
&                & = Q_{\Lambda_1}(\beta \textbf{y}_R + \textbf{t}) \mod\Lambda_2. \label{vrc}
\end{eqnarray}
It can be seen easily that $\textbf{v}_{Rr} \in \mathcal{C}_r$ and $\textbf{v}_{Rc} \in \mathcal{C}_c$.
We obtain the same common information generated in \eqref{vr}. Thus, the (total) information that is intended to terminal T1 is such that 
\begin{subequations}
\begin{align}
\mathbf{v}_{R1} & = \mathbf{v}_{Rr} + \mathbf{v}_{Rc} \label{vr1} \\
                & = Q_{\Lambda_0}(\beta \mathbf{y}_R + \mathbf{t})\mod\Lambda_1 +  Q_{\Lambda_1}(\beta \mathbf{y}_R + \mathbf{t}) \mod\Lambda_2 \label{lp1}\\
                & = Q_{\Lambda_0}(\beta \mathbf{y}_R + \mathbf{t}) - Q_{\Lambda_1}(Q_{\Lambda_0}(\beta \mathbf{y}_R + \mathbf{t})) + Q_{\Lambda_1}(\beta \mathbf{y}_R + \mathbf{t}) -  Q_{\Lambda_2}(Q_{\Lambda_1}(\beta \mathbf{y}_R + \mathbf{t})) \label{lp2}\\
                & = Q_{\Lambda_0}(\beta \mathbf{y}_R + \mathbf{t}) - Q_{\Lambda_2}(\beta \mathbf{y}_R + \mathbf{t}) \label{lp3}\\
                & = Q_{\Lambda_0}(\beta \mathbf{y}_R + \mathbf{t}) - Q_{\Lambda_2}( Q_{\Lambda_0}(\beta \mathbf{y}_R + \textbf{t})) \label{lp4} \\
                & = Q_{\Lambda_0}(\beta \mathbf{y}_R + \mathbf{t}) \mod\Lambda_2.
\end{align}           
\end{subequations}
where the Eq. \eqref{lp2}, \eqref{lp3} and \eqref{lp4} follow using the properties of the modulo operation as given in Section \ref{SubSec::lcodes}.  
\subsubsection{Decoding}
$\textbf{v}_{Rc}$ is decoded at terminal T$_2$. Then, $\hat{\textbf{u}}_{2}$ is reconstructed with an LSD using the side information $\textbf{s}_2$ as 
\begin{equation}
\label{ur2}
\hat{\textbf{u}}_2 = \gamma_2((\textbf{v}_{Rc} -\textbf{t} - \beta \textbf{s}_2) \mod\Lambda_2).
\end{equation} 

\begin{figure}[htbp]
\centering
\includegraphics[width =0.7\linewidth]{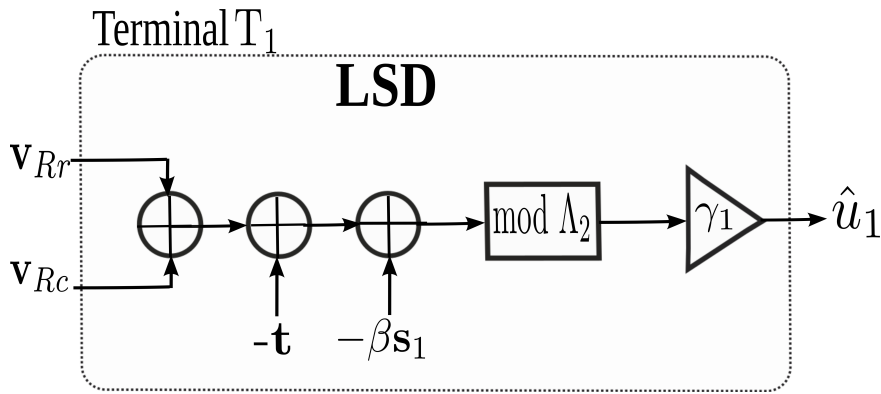}
\caption{Lattice source decoding at the Terminal 1}\label{Fig4}
\end{figure}

At terminal T$_1$, $\textbf{v}_{Rc}$ and $\textbf{v}_{Rr}$ are both decoded correctly.   
These coset leaders are used to recalculate the total information $\textbf{v}_{R1}$ from \eqref{vr1}. Finally, the decoder reconstructs $\hat{\textbf{u}}_1$ as defined by \eqref{ur1} and shown in Fig. \ref{Fig4}, as
\begin{equation}
\label{ur1}
\hat{\textbf{u}}_1 = \gamma_1((\textbf{v}_{R1} -\textbf{t} - \beta \textbf{s}_1) \mod\Lambda_2)
\end{equation}
\subsection{Rate Analysis}
\label{SubSec2::rateAnalysis}
At the relay, the relay generates the indices of $\textbf{v}_{Rc}$ and $\textbf{v}_{Rr}$. Then they are mapped to the channel codewords $\textbf{x}_{Rc}$ and $\textbf{x}_{Rr}$.
The relay sends $\textbf{x}_{R}(m_R)$ which is the superposition of $\textbf{x}_{Rc}$ and $\textbf{x}_{Rr}$ with transmit power $\nu P_R$ and $(1-\nu) P_R$, $\nu \in \{0,1\}$, respectively.
The refinement codeword $\textbf{x}_{Rr}$ is encoded on top of the common codeword $\textbf{x}_{Rc}$ and it is treated as an interference while decoding the common message. Thus, $\textbf{X}_{Rc} \rightarrow \textbf{X}_{r} \rightarrow(\textbf{Y}_1, \textbf{Y}_2)$ forms a Markov chain.
As described in previous one layer PNC scheme, the broadcast rate is bounded by the worst relay-terminal channel capacity for the common message, and by the relay-T$_1$ channel for the refinement message. In addition, the source-channel separation ensures that the codewords $\textbf{x}_{Rc}$ and $\textbf{x}_{Rr}$ are transmitted reliably to the terminals and that $\textbf{v}_{Rc}$ and $\textbf{v}_{Rr}$ are available at the LSD input of corresponding receivers. Therefore, the rates are such that

\begin{align}
n_1 R_{c} & \leq n_2 \min\{ I(X_{Rc}; Y_1), I(X_{Rc}; Y_2)\} \label{SCcst1} \\
n_1 R_{r} &\leq n_2 I(X_{Rr}; Y_1 |X_{Rc})\label{SCcst2} 
\end{align}
For real Gaussian codebooks, we have
$$I(X_{Rc}; Y_1) = \frac{1}{2} \log_2\left(1 + \displaystyle \frac{\nu |h_1|^2 P_R}{(1- \nu) |h_1|^2 P_R + \sigma_1^2}\right)$$
$$I(X_{Rc}; Y_2) = \frac{1}{2} \log_2\left(1 + \displaystyle\frac{\nu |h_2|^2 P_R}{(1- \nu) |h_2|^2 P_R + \sigma_2^2}\right)$$
$$I(X_{Rr}; Y_1 |X_{Rc}) = \frac{1}{2} \log_2\left(1 + \displaystyle\frac{(1-\nu)|h_1|^2 P_R}{\sigma_1^2}\right)$$
Since $|h_2| \leq |h_1|$, $\min\{ I(X_{Rc}; Y_1), I(X_{Rc}; Y_2)\} = I(X_{Rc}; Y_2)$. Using Eq. \eqref{Rcref}, \eqref{Rrref}, \eqref{SCcst1} and \eqref{SCcst2}, the rates' conditions become 
\begin{eqnarray}
\label{ConRref} 
&n_1 \log_2\left(\displaystyle\frac{\sigma^2(\Lambda_2)}{\sigma^2(\Lambda_1)}\right) &\leq n_2 \log_2\left(1 + \displaystyle\frac{\nu |h_2|^2 P_R}{(1- \nu) |h_2|^2 P_R + \sigma_2^2}\right) \\
&n_1 \log_2\left(\displaystyle \frac{\sigma^2(\Lambda_1)}{\sigma^2(\Lambda_0)} \right) &\leq
n_2 \log_2\left(1 + \displaystyle\frac{(1-\nu)|h_1|^2 P_R}{\sigma_1^2}\right) \label{ConRref2} 
\end{eqnarray}
Now, $\hat{\textbf{u}}_1$ and $\hat{\textbf{u}}_2$ can be obtained using \eqref{ur1} and \eqref{ur2}, respectively.
At terminal T$_2$, $\hat{\textbf{u}}_2$ can be written as:
\begin{eqnarray}
&\hat{\textbf{u}}_2 & = \gamma_2((\beta \textbf{u}_2 + \textbf{e}_{q,1}) \mod\Lambda_2) \label{hatu2}\\
& & = \gamma_2(\beta \textbf{u}_2 + \textbf{e}_{q,1}) \label{hatu2eq}
\end{eqnarray}
where $\textbf{e}_{q,1}$ is the quantization error at lattice $\Lambda_1$ given by
$$\textbf{e}_{q,1} = Q_{\Lambda_1}\left(\beta \textbf{y}_R + \textbf{t}) \right) - (\beta \textbf{y}_R + \textbf{t}) = -(\beta \textbf{y}_R + \textbf{t}) \mod \Lambda_1$$
and \eqref{hatu2eq} can be obtained by proceeding as in Section \ref{Subsec::rateAnalysis}.
Note that $\text{Pr}(\beta
\mathbf{U}_2 + \mathbf{E}_{q,1} \notin \mathcal{V}_{2})$ vanishes asymptotically provided that:
\begin{equation}
\frac{1}{n_1} \mathbb{E}\|\beta \mathbf{U}_2  + \mathbf{E}_{q,1} \|^2 = \beta^2 \sigma^2_{U_2} + \sigma^2(\Lambda_1) \leq \sigma^2(\Lambda_2)
\label{consSref}
\end{equation}
In this case, the rate achievable at terminal T$_2$ is such that 
\begin{equation}
\label{R12reftmp}
n R_{12} \leq \frac{n_1}{2} \log_2\left(1 + \frac{\beta^2 |h_1|^2 P_1}{\beta^2 \sigma_R^2 + \sigma^2(\Lambda_1)}\right).
\end{equation}
At terminal T$_1$, $\hat{\textbf{u}}_1$ can be obtained as
\begin{eqnarray}
&\hat{\textbf{u}}_1 & = \gamma_1((\beta \textbf{u}_1 + \textbf{e}_{q,0}) \mod\Lambda_2) \label{hatu1}\\
& & \equiv \gamma_1(\beta \textbf{u}_1 + \textbf{e}_{q,0})\label{hatu1eq}
\end{eqnarray}
where $\textbf{e}_{q,0}$ is the modulo-$\Lambda_0$ quantization error given by
$$\textbf{e}_{q,0} = Q_{\Lambda_0}(\beta \textbf{y}_R + \textbf{t}) - (\beta \textbf{y}_R + \textbf{t}) = -(\beta \textbf{y}_R + \textbf{t}) \mod \Lambda_0$$
and \eqref{hatu1eq} holds if $\beta \textbf{u}_1 +\textbf{e}_{q,0} \in \mathcal{V}_2$. Note that, by using Lemma 1,
$\textbf{E}_{q,0}$ is independent from $\textbf{Y}_R$, and thus from $\textbf{U}_1$. Also this quantization error is uniformly distributed over $\mathcal{V}_0$. Therefore, $\text{VAR}(E_{q,0}) = \sigma^2(\Lambda_0)$. The probability $\text{Pr}(\beta
\mathbf{U}_1  + \mathbf{E}_{q,0} \notin \mathcal{V}_{2})$ vanishes asymptotically provided that:
\begin{equation}
\frac{1}{n_1} \mathbb{E}\|\beta \mathbf{U}_1  + \mathbf{E}_{q,0} \|^2 = \beta^2 \sigma^2_{U_1} + \sigma^2(\Lambda_0) \leq \sigma^2(\Lambda_2)
\label{consSref2}
\end{equation}
Thus, $$\hat{\textbf{U}}_1 = \gamma_1(\beta h_1 \textbf{X}_2 + \beta \textbf{Z}_R + \textbf{E}_{q,0}) $$
Communication from terminal T$_2$ to terminal T$_1$ is equivalent to that over an AWGN channel with noise $\gamma_1(\beta \textbf{Z}_R + \textbf{E}_{q,0})$.
Hence the achievable rate of this link satisfies:
\begin{equation}
\label{R21reftmp}
n R_{21} \leq \frac{n_1}{2} \log_2\left(1 + \frac{\beta^2 |h_2|^2 P_2}{\beta^2 \sigma_R^2 + \sigma^2(\Lambda_0)}\right)
\end{equation}
\subsection{Achievable Rate Region}
\label{Subsec2::achievblRates}
The rate region that is achievable using the coding scheme that we described so far can be obtained using \eqref{ConRref},\eqref{ConRref2}, \eqref{consSref} and \eqref{consSref2}.
Letting $\displaystyle \frac{n_1}{n} = \alpha$, we get
\begin{align*}
\begin{cases}
\displaystyle \frac{\sigma^2(\Lambda_2)}{\sigma^2(\Lambda_1)}\leq \left(1+ \displaystyle \frac{\nu |h_2|^2 P_R}{(1 - \nu) |h_2|^2 P_R 
+ \sigma_2^2}\right)^{\frac{1 -\alpha}{\alpha}} \\
\displaystyle \frac{\sigma^2(\Lambda_1)}{\sigma^2(\Lambda_0)}\leq \left(1+ \displaystyle \frac{(1 -\nu) |h_1|^2 P_R}{\sigma_1^2}\right)^{\frac{1 -\alpha}{\alpha}}\\
\sigma^2(\Lambda_1) \leq \sigma^2(\Lambda_2) - \beta^2 \sigma^2_{U_2} \\
\sigma^2(\Lambda_0) \leq \sigma^2(\Lambda_2) - \beta^2 \sigma^2_{U_1} 
\end{cases}
\end{align*}
Since $\sigma^2(\Lambda_2) \geq \sigma^2(\Lambda_1) \geq \sigma^2(\Lambda_0)$, the last constraint in the system is not active. Thus we obtain the following bounds on the second moment of the lattices
\begin{align}
\label{LBsigma1ref}
\sigma^2(\Lambda_1) &\geq \frac{\beta^2 \sigma^2_{U_2}}{\left(1+ \displaystyle \frac{\nu |h_2|^2 P_R}{(1 - \nu) |h_2|^2 P_R 
+ \sigma_2^2}\right)^{\frac{1 -\alpha}{\alpha}} -1} \\
\sigma^2(\Lambda_0) &\geq \frac{\sigma^2_{\Lambda_1}}{\left(1+ \displaystyle \frac{(1 -\nu) |h_1|^2 P_R}{\sigma_1^2}\right)^{\frac{1 -\alpha}{\alpha}}} \label{LBsigma0ref}
\end{align}
The rate region defined by \eqref{R12reftmp} and \eqref{R21reftmp} can then be rewritten equivalently as
\begin{eqnarray}
R_{12} \leq \frac{\alpha}{2} \log_2\left(1 + \mbox{SNR}_{1 \rightarrow 2} \right)\\
R_{21} \leq  \frac{\alpha}{2} \log_2\left(1 +\mbox{SNR}_{2 \rightarrow 1} \right)
\end{eqnarray}
where the end-to-end SNRs are given by
\begin{eqnarray}
\label{SNR12ref}
\mbox{SNR}_{1 \rightarrow 2} = \frac{\beta^2 |h_1|^2 P_1}{\beta^2 \sigma_R^2 + \sigma^2(\Lambda_1)} \\
\mbox{SNR}_{2 \rightarrow 1} = \frac{\beta^2 |h_2|^2 P_2}{\beta^2 \sigma_R^2 + \sigma^2(\Lambda_0)} \label{SNR21ref}
\end{eqnarray}

It is easy to see that one obtains larger rates if the inequalities in (55) and (56) hold with equality, i.e., the optimal choice on the second moment of $\Lambda_1$ is
\begin{equation}
\label{sigmaLambda1}
\sigma^2(\Lambda_1)_{\min} = \frac{\beta^2 \sigma^2_{U_2}}{\left(1+ \displaystyle \frac{\nu |h_2|^2 P_R}{(1 - \nu) |h_2|^2 P_R 
+ \sigma_2^2}\right)^{\frac{1 -\alpha}{\alpha}} -1}
\end{equation}
and the optimal choice on the second moment of $\Lambda_0$ is
\begin{equation}
\label{sigmaLambda0}
\sigma^2(\Lambda_0)_{\min} = \displaystyle  \frac{\beta^2 \sigma^2_{U_2}}{\left(1 + \frac{(1 -\nu) |h_1|^2 P_R}{ \sigma_1^2}\right)^{\frac{1- \alpha}{\alpha}} \left[ \left(1+ \frac{\nu |h_2|^2 P_R}{(1 -\nu) |h_2|^2 P_R + \sigma_2^2}\right)^{\frac{1 - \alpha}{\alpha}} -1 \right]}
\end{equation}
Finally, by substituting  $\sigma^2(\Lambda_1)_{\min}$ and $\sigma^2(\Lambda_0)_{\min}$ in \eqref{SNR12ref} and \eqref{SNR21ref}, we get \eqref{R12ref} and \eqref{R21ref}. This completes the proof of Theorem 2.
\begin{remark}
The obtained achievable rates are independent of the choice of the scaling factors $\beta$ and $\gamma_i$. The optimal choice of these parameters is explained when considering the source coding problem as explained in the next section.
\end{remark}
\subsection{Analog Signal Transmission}
\label{Subsec::distrref}
Proceeding as in the analysis in \ref{Subsec::distr}, it can be easily obtained that the optimal scaling factors $\gamma_i$ that minimize the distortion at each terminal are given by
\begin{eqnarray}
\label{gamma1ref}
\gamma_1^* & = \frac{\beta \sigma^2(\Lambda_1) }{\beta^2 \sigma^2_{U_2} + \sigma^2(\Lambda_1) },\\
\gamma_2^* & = \frac{\beta \sigma^2(\Lambda_0) }{\beta^2 \sigma^2_{U_1} + \sigma^2(\Lambda_0) }. \label{gamma2ref}
\end{eqnarray}
Thus, the minimal distortion at terminal T$_2$ is 
\begin{equation}
D_2^{\min} = \frac{\sigma^2_{U_2}}{\left(1+ \displaystyle \frac{\nu |h_2|^2 P_R}{(1 - \nu) |h_2|^2 P_R 
+ \sigma_2^2}\right)^{\frac{1 -\alpha}{\alpha}}}
\end{equation}
and the minimal distortion at terminal T$_1$ is
\begin{eqnarray}
&D_1^{\min} & = \frac{\sigma^2_{U_1} \sigma^2(\Lambda_0)_{\min}  }{\beta^2 \sigma^2_{U_1} + \sigma^2(\Lambda_0)_{\min}} \\
&            & =  \frac{\sigma^2_{U_2} \sigma^2_{U_1} }{\left(1+ \displaystyle \frac{(1 -\nu) |h_1|^2 P_R}{\sigma_1^2}\right)^{\frac{1 -\alpha}{\alpha}}\left(\left(1+ \displaystyle \frac{\nu |h_2|^2 P_R}{(1 - \nu) |h_2|^2 P_R 
+ \sigma_2^2}\right)^{\frac{1 -\alpha}{\alpha}}- 1\right) \sigma^2_{U_1} + \sigma^2_{U_2}}.
\end{eqnarray}
Observe that the distortion $D_1^{\min}$ that is allowed by the layered coding scheme described so far is, as expected, smaller than that of the coding scheme of Section III given by \eqref{Dimin}.

To summarize, if we are interested in the distortion problem in addition to the transmission problem addressed in this paper, the choice of $\beta$ can be left to the designer. The optimal lattice parameters and the receivers' scaling factors that depend on this choice are given by Eq. \eqref{sigmaMim} and \eqref{gammaistar} for the first scheme and \eqref{sigmaLambda1}, \eqref{sigmaLambda0}, \eqref{gamma1ref} and \eqref{gamma2ref} for the second scheme. However, this choice does not affect the optimal achievable rates and distortions that depend only on the system parameters.
\section{Numerical Results}
\label{Sec::sims}
This section presents numerical results of the achievable rates of our proposed schemes compared to AF and DF protocols and the outer-bound capacity given in \cite{gdrisis.k07, sarnoff.kdmt08}.

We select the time-division parameter $\alpha \in [0, 1]$ that permits to trade among the multiaccess and broadcast phases in a manner that maximizes the users rates.
The bounds are determined by maximizing the weighted sum of the rates $R_{12}$ and $R_{21}$ for each protocol. For example, for the scheme of Section \ref{Sec::twolayers}, we solve the following problem for all values of $\eta \in [0,1]$
\begin{subequations}
\begin{eqnarray}
\label{optiProb}
\max & \eta R_{12} + (1 - \eta) R_{21}\\
\mbox{s.t.}& ( R_{12}, R_{21}) ~\mbox{satisfy}~  \eqref{R12ref} ~\mbox{and}~ \eqref{R21ref} \\
 & ~\mbox{for}~ \alpha ~\mbox{and}~ \nu \in [0, 1] 
\end{eqnarray}
\end{subequations}
It is worth noting that the time division $\alpha$ with AF relaying scheme is is set optimally to $\frac{1}{2}$. 

We consider equal noise variances $\sigma^2_1$ = $\sigma^2_2$ =$\sigma^2_R$ = 1, different transmit powers and asymmetric channels with $|h_1|^2 P_1 \geq |h_2|^2 P_2$. For convenience, we refer to the achievable rate regions of Theorems \textsl{\ref{Theorem1}} and \textsl{\ref{Theorem2}} respectively as LCF1 and LCF2.

\begin{figure}
    \begin{subfigure}[b]{0.5\textwidth}{
       \includegraphics[width=1\linewidth]{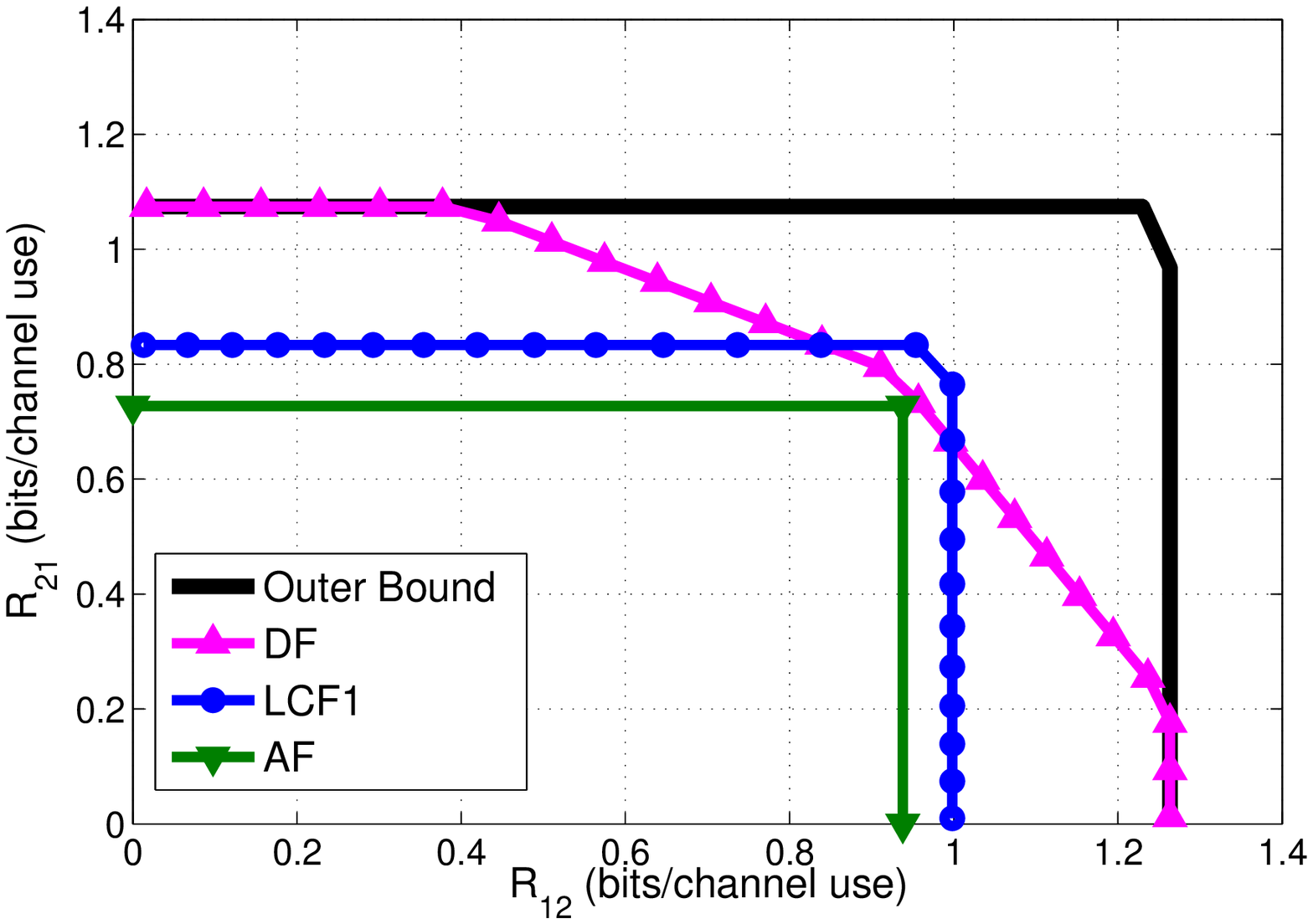}
       \caption{P$_1$ = 15 dB, P$_2$ = 10 dB, P$_R$ = 20 dB, \\ $|h_1|^2 = 0.5$, $|h_2|^2 = 1$}
       \label{SimFig1a}}
    \end{subfigure}
    \begin{subfigure}[b]{0.5\textwidth}{
       \includegraphics[width=1\linewidth]{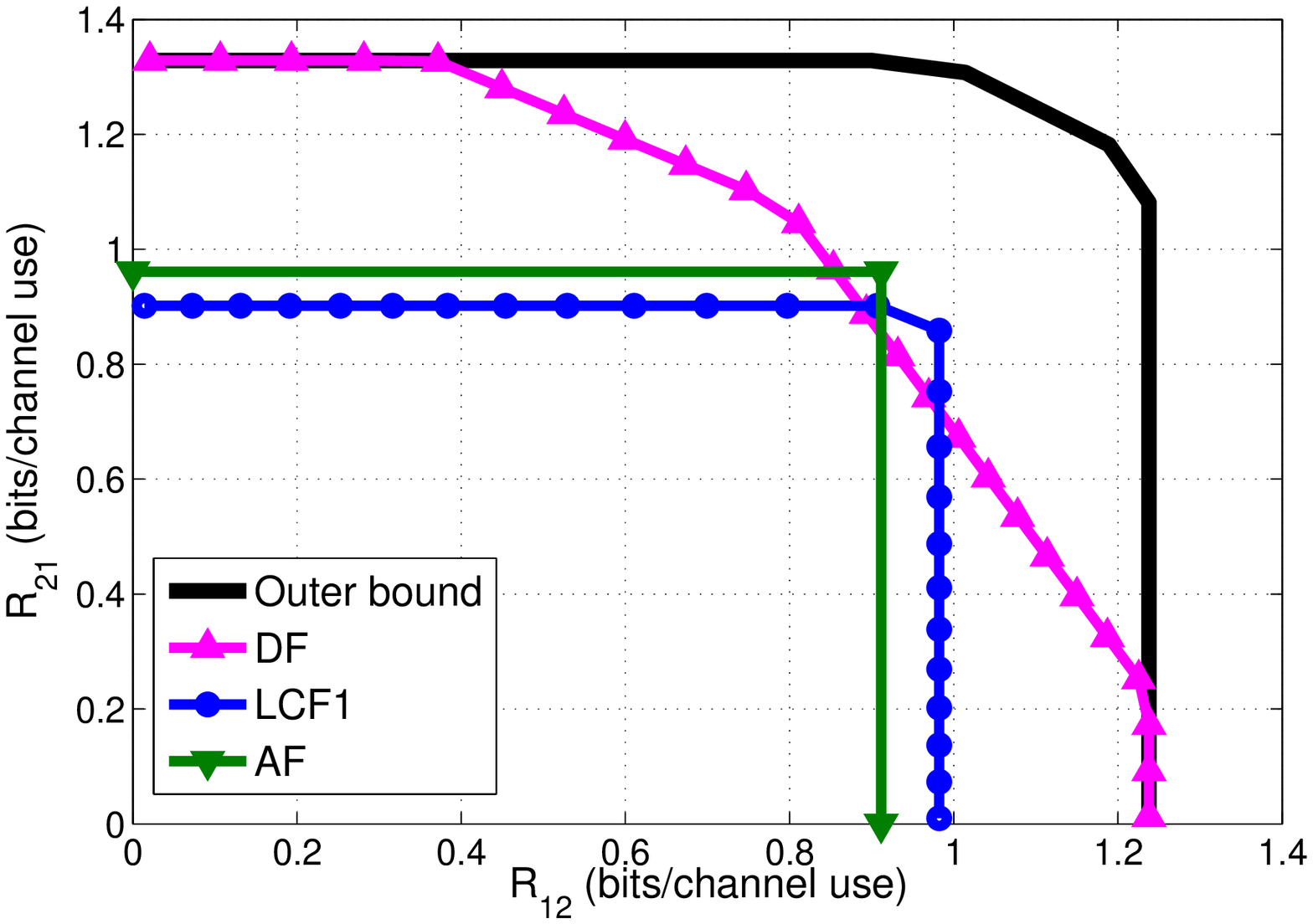}
       \caption{P$_1$ = 10 dB, P$_2$ = 15 dB, P$_R$ = 20 dB, \\ $|h_1|^2 = 2$, $|h_2|^2 = 0.5$}
       \label{SimFig1b}}
       \end{subfigure}
      \caption{Achievable rate regions and the outer bound capacity of the Gaussian TWRC. In the left, T$_1$ has the best transmit power and the worst channel. In the right, T$_2$ has the best transmit power and the worst channel.} 
         \label{SimFig1}
\end{figure}

Figure \ref{SimFig1} shows the rates allowed by AF, DF and our proposed scheme LCF1 for two different setups: i) terminal T2 experiencing better channel conditions and having less power than terminal T$_1$ in Fig. \ref{SimFig1a}, and ii) terminal T$_1$ experiencing better channel conditions and having less power than terminal T$_2$ in Fig. \ref{SimFig1b}. 

Note that our scheme LCF1 is, in essence, a CF relaying strategy that is adopted and tailored appropriately for the TWRC. Being based on linear (lattice) coding, this strategy has been shown in \cite{atc.skszd12} to possibly achieve the same rates as those allowed by random coding \cite{sarnoff.kdmt08, allerton.gtn08}.
It has been shown in \cite{sarnoff.kdmt08}, that CF strategy achieves rates that are larger than those by AF for symmetric power and channel configurations.
However, this result is not verified for asymmetric channels. This is shown in Fig.\ref{SimFig1} where the difference between the rate regions of AF and LCF1 is negligible for moderate SNR values and asymmetric channels. 

\begin{figure}
\centering
       \includegraphics[width=0.5\linewidth]{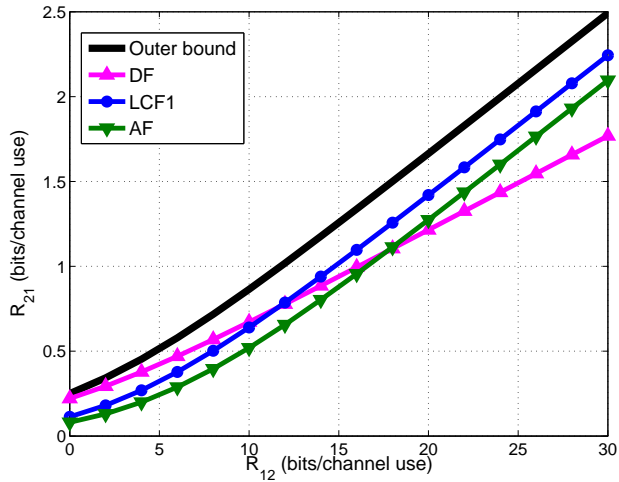}
       \caption{Equal rates $R_{12} = R_{21}$ for symmetric channels: SNR = SNR$_{1R}$ = SNR$_{R1}$ = SNR$_{2R}$ =SNR$_{R2}$. LCF1 outperforms AF and DF for SNR $>$ 11 dB }
       \label{SimFig2}
\end{figure}

Figure \ref{SimFig2} illustrates the performance of all schemes in the symmetric power and channel conditions case. End-to-end equal rates $R_{12} =R_{21}$ as a function of the SNR are shown for
equal channel and power conditions for all nodes. Define SNR$_{ij} = \frac{|h_{ij}|^2 P_i}{\sigma^2_j}$. It is clearly seen that LCF1 outperforms DF for SNRs $\geq$ 12 dB. 
This result can be interpreted analytically. In fact, it can be seen easily that for small SNR values, DF rate approaches
$$
R_{DF} \rightarrow \max_{\alpha} \min \{\alpha SNR, (1 -\alpha) SNR \} = \frac{1}{4} SNR.
$$
Also, the rate offered by LCF1 approaches
$$
R_{LCF1} \rightarrow \frac{((\sqrt{SNR+1} -1) + (SNR -2 \sqrt{SNR} + 2) \sqrt{SNR})SNR^2}{2(\sqrt{SNR + 1} -1) +\sqrt{SNR}}
$$
Thus, in such small SNR regime, we have $R_{LCF1} \leq R_{DF}$. On the other hand, for high SNR, DF rate can be approximated by 
$$
R_{DF} \rightarrow \frac{1}{6} \log_2(SNR)
$$
and LCF1 rate approaches 
$$
R_{LCF1} \rightarrow \frac{1}{4} (\log_2(SNR) -1).
$$
It is immediately seen that for large SNRs, we have, $R_{LCF1} \geq R_{DF}$ which corresponds to the result in Fig. \ref{SimFig2}.

\begin{figure}
   \begin{subfigure}[b]{0.5\textwidth}{
       \includegraphics[width=1\linewidth]{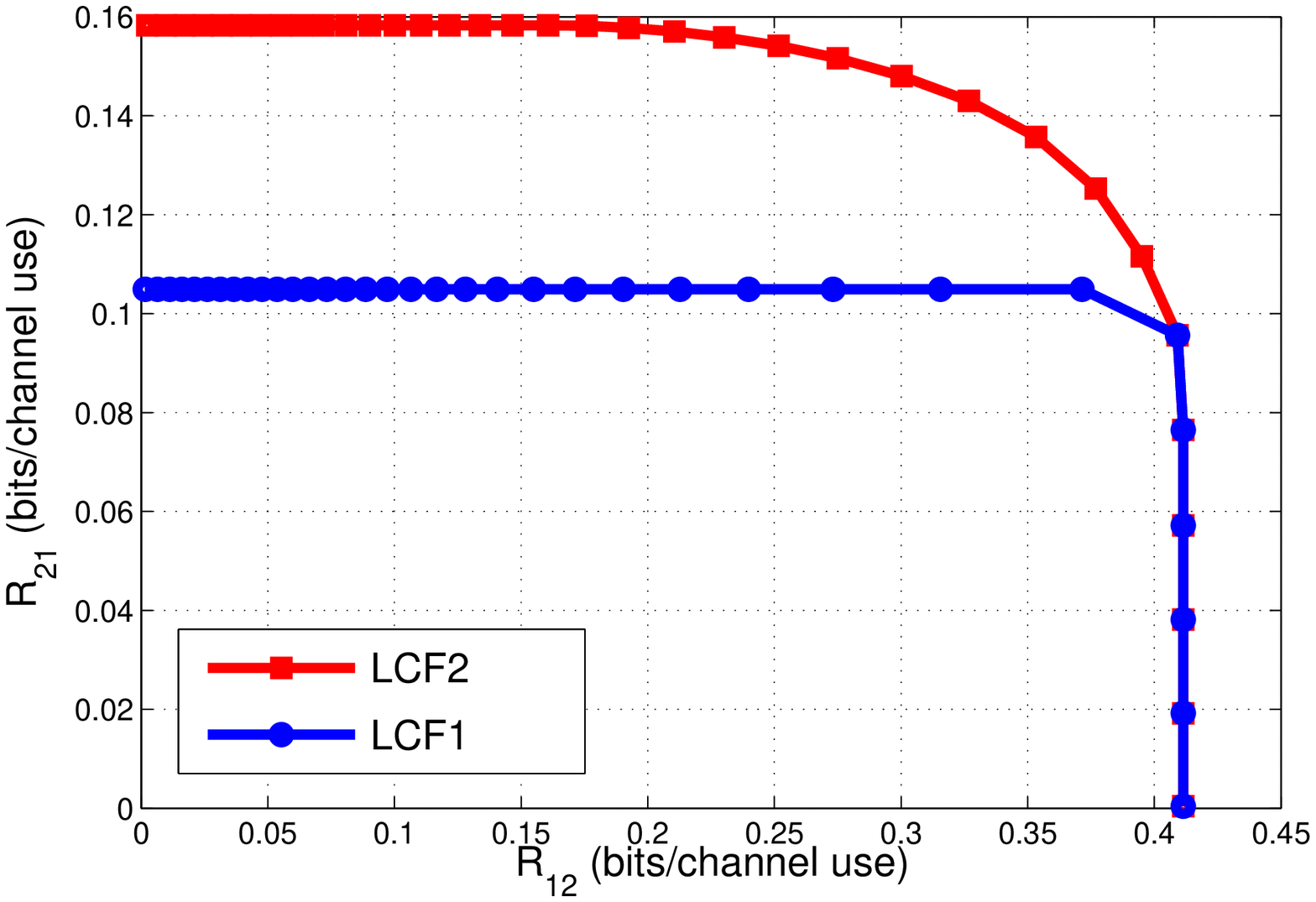}
       \caption{P$_1$ = 10 dB, P$_2$ = P$_R$ = 5 dB, $|h_1|^2 = 2$, $|h_2|^2 = 0.5$}
       \label{SimFig3a}}
    \end{subfigure}
        \begin{subfigure}[b]{0.5\textwidth}{
       \includegraphics[width=1\linewidth]{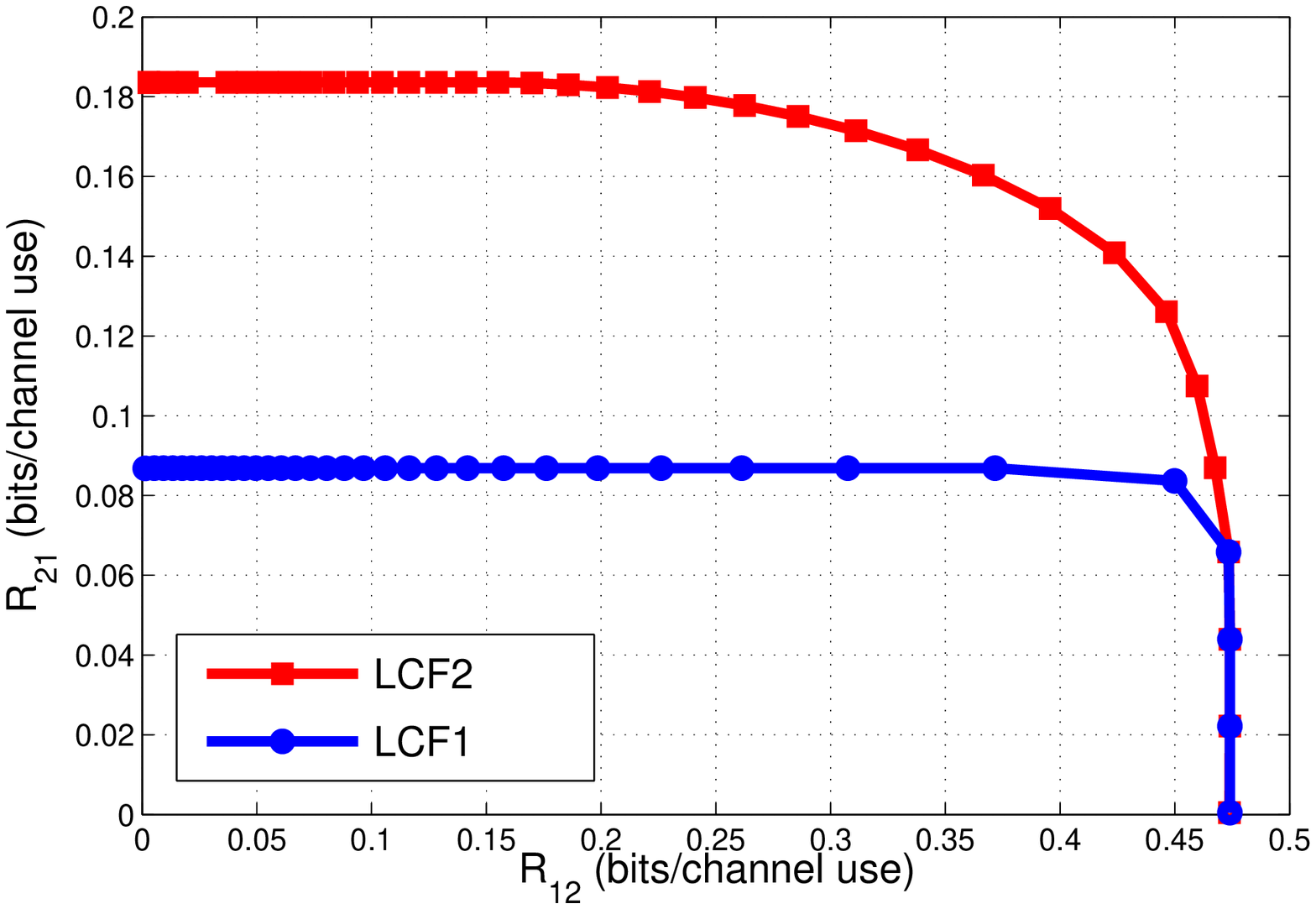}
        \caption{P$_1$ = 10 dB, P$_2$ = P$_R$ = 5 dB, $|h_1|^2 = 6$, $|h_2|^2 = 0.5$}
       \label{SimFig3b}}
       \end{subfigure}
       \caption{Achievable rate regions of LCF1 and LCF2. LCF2 achieves greater end-to-end rates at T$_1$}
       \label{SimFig3}
\end{figure}

In what follows, we consider channel parameters combinations such that $P_1 \geq P_2$ and $|h_1|^2 \geq |h_2|^2$. Figure \ref{SimFig3} draws the achievable rate regions of LCF1 and LCF2. One can see that the two-layer based scheme LCF2 enlarges the rate region compared to the basic scheme since the relay sends additional information to the best terminal T$_1$. For the setting presented in Fig. \ref{SimFig3a}, the achievable rate $R_{21}$ increases by $60\% $ due to the additional refinement individual description.
Figure  \ref{SimFig3b} illustrates this aspect for a different choice of the channel parameters where $R_{21}$ increases by more than $100\% $.

Finally, when compared to DF and AF relaying schemes, simulations show that LCF2 scheme outperforms AF in all SNR regimes for symmetric and asymmetric configurations. 

\begin{figure}
    \begin{subfigure}[b]{0.5\textwidth}{
       \includegraphics[width=1\linewidth]{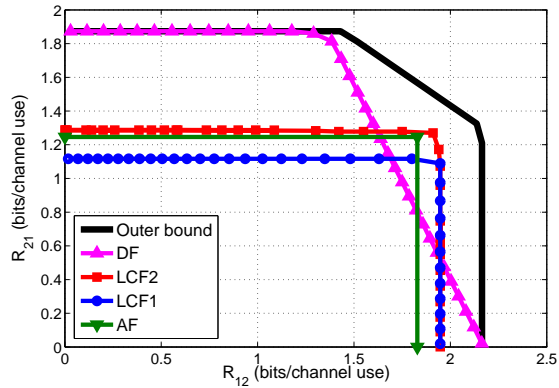}
        \caption{P$_1$ = 30 dB, P$_2$ = 25 dB, P$_R$ = 30 dB, $|h_1|^2 = 1$, $|h_2|^2 = 0.2$}
       \label{SimFig4a}}
              \end{subfigure}
                  \hspace*{1mm}
    \begin{subfigure}[b]{0.5\textwidth}{
       \includegraphics[width=1\linewidth]{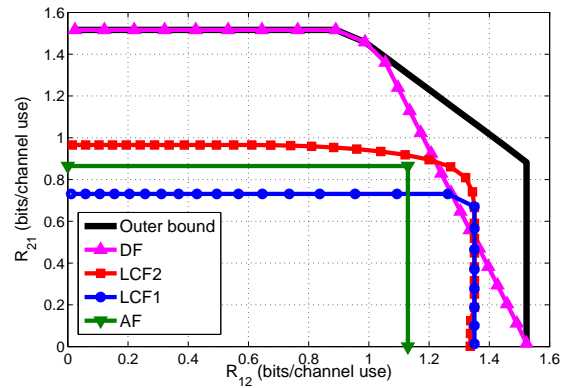}
       \caption{P$_1$ = 20 dB, P$_2$ = 18 dB, P$_R$ = 17 dB, $|h_1|^2 = 4$, $|h_2|^2= 0.5$}
       \label{SimFig4b}}
       \end{subfigure}
    \begin{subfigure}[b]{0.5\textwidth}{
       \includegraphics[width=1\linewidth]{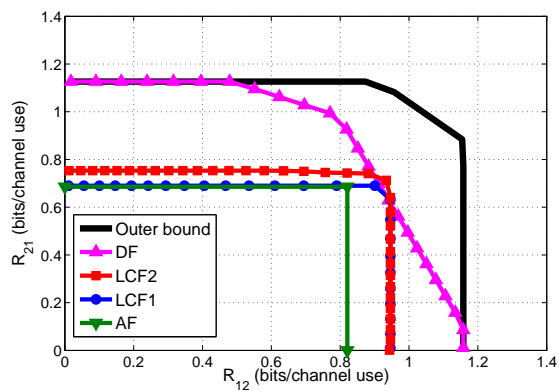}
        \caption{P$_1$ = 10 dB, P$_2$ = 9 dB, P$_R$ = 9 dB, $|h_1|^2 = 4$ and $|h_2|^2 = 2$}
       \label{SimFig4c}}
       \end{subfigure}
                  \hspace*{1mm}
   \begin{subfigure}[b]{0.5\textwidth}{
       \includegraphics[width=1\linewidth]{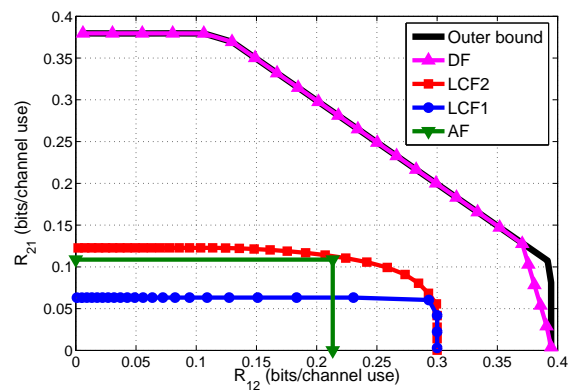}
        \caption{P$_1$ = 5 dB, P$_2$ = 3 dB, P$_R$ = 3 dB, $|h_1|^2 = 4$ and $|h_2|^2 = 0.5$}
       \label{SimFig4d}}
       \end{subfigure}
       \caption{Achievable rate regions of DF, AF, LCF1 and LCF2 in different channel and power settings}
       \label{SimFig4}
\end{figure}

Figure \ref{SimFig4} illustrates the achievable rate regions of DF, AF and both lattice-based schemes, LCF1 and LCF2, for various SNR settings.

At small SNRs, the scheme LCF2 outperforms the scheme LCF1; but they both fall short of attaining the same performance as that offered by DF which is nearly optimal in this SNR regime. 
In fact, in this SNR regime, the rate region obtained with DF relaying approaches relatively closely the outer bound as can be seen in Fig. \ref{SimFig4d}. 
Note that our observation here is consistent with
the results in 
\cite{sarnoff.kdmt08, ieeeIT.kdmt11} that showed that DF scheme is better than the other relaying schemes for low SNR region. 

At very large SNRs, LCF1 and LCF2 achieve better rates than DF as shown in Fig. \ref{SimFig4a}. At moderate to large SNRs, the scheme LCF2 performs better than classic DF.

\begin{remark}
\label{remarkFin}
We have assumed in our system model perfect CSI at all nodes. However, in the proposed two lattice-based coding schemes (LCF1 and LCF2), this perfect knowledge of the channel state can be relaxed. In fact, in order to compress its received signal, the relay needs only the module of the channel gains to reconstruct its encoding scheme. For each terminal, the decoder uses the available side information $\textbf{S}_i = h_i \textbf{X}_i$ that depends on its terminal-relay channel.
Appropriate training sequences can be employed to estimate the channel of the relay.
Furthermore, each decoder estimates only its unknown part of the relay received signals. It is shown in sections \ref{Subsec::rateAnalysis} and \ref{SubSec2::rateAnalysis} that the communication between both terminals is equivalent to the output of an effective Gaussian channel
for both proposed schemes. Thus, a training sequence can also be used in order to estimate at each decoder, the channel on the other link.
\end{remark}
\section{Conclusion}
\label{Sec::concl}
In this paper, we studied the problem of exchanging messages over a Gaussian two-way relay channel. We derived two achievable rate regions based on compress and forward lattice coding.
In the proposed schemes, the relay uses a lattice based Wyner-Ziv encoding by taking into account the presence of the side information at each node. (i.e. the signal broadcasted by the relay includes also the signal that has been transmitted by each user to the relay during the first MAC transmission phase). 

First, we develop a coding scheme in which the relay broadcasts the same signal to both terminals. 
We show that this scheme offers the same performance as random coding based compress-and-forward protocol \cite{atc.skszd12}.
Then, we propose, and analyze the performance of, an improved coding scheme in which the relay sends not only a common description of its output, but also an individual description that is destined to be recovered by only the user who experiences better channel conditions and better side information. We show that this results in substantial gains in rates.
Numerical results demonstrate an enhancement of the achievable rate region 
over the basic scheme up to 100$\%$ for moderate SNR regime and asymmetric channel conditions. 
Also, the improved scheme outperforms classic amplify-and-forward at all SNR values, and classic decode-and-forward for certain SNR regimes.

Finally, it is worth mentioning that our schemes are based on structured codes that have low complexity compared to random coding from practical viewpoints. However, in these schemes, lattices codewords are used only at the relay while Gaussian codewords are used at the terminals' nodes. Considering lattice codes at all the nodes can be even more appropriate for practical systems.

\bibliographystyle{IEEEtran}
\bibliography{myalienBiblio} 

\begin{thebibliography}{10}
\providecommand{\url}[1]{#1}
\csname url@samestyle\endcsname
\providecommand{\newblock}{\relax}
\providecommand{\bibinfo}[2]{#2}
\providecommand{\BIBentrySTDinterwordspacing}{\spaceskip=0pt\relax}
\providecommand{\BIBentryALTinterwordstretchfactor}{4}
\providecommand{\BIBentryALTinterwordspacing}{\spaceskip=\fontdimen2\font plus
\BIBentryALTinterwordstretchfactor\fontdimen3\font minus
  \fontdimen4\font\relax}
\providecommand{\BIBforeignlanguage}[2]{{%
\expandafter\ifx\csname l@#1\endcsname\relax
\typeout{** WARNING: IEEEtran.bst: No hyphenation pattern has been}%
\typeout{** loaded for the language `#1'. Using the pattern for}%
\typeout{** the default language instead.}%
\else
\language=\csname l@#1\endcsname
\fi
#2}}
\providecommand{\BIBdecl}{\relax}
\BIBdecl

\bibitem{ieeeIT.acly00}
R.~Ahlswede, N.~Cai, S.-Y.~R. Li, and R.~W. Yeung, ``Network information
  flow,'' \emph{IEEE Trans. on Inform. Theory}, vol.~46, no.~4, pp. 1204--1216,
  Jul. 2000.

\bibitem{mobicom.zll06}
S.~Zhang, S.~Liew, and P.~Lam, ``Physical layer network coding,'' in \emph{ACM
  MOBICOM}, Los Angeles, USA, 2006.

\bibitem{gdrisis.k07}
R.~Knopp, ``Two-way wireless communication via a relay station,'' in
  \emph{GDR-ISIS meeting}, Paris, France, Mar. 2007.

\bibitem{ieeeIT.kmt08}
S.~J. Kim, P.~Mitran, and V.~Tarokh, ``Performance bounds for bi-directional
  coded cooperation protocols,'' \emph{IEEE Trans. on Inform. Theory}, vol.~54,
  no.~11, pp. 5235--5241, Nov. 2008.

\bibitem{acssc.rw05}
B.~Rankov and A.~Wittneben, ``Spectral efficient signaling for half-duplex
  relay channels,'' in \emph{Asilomar Conference on Signals, Systems and
  Computers (ACSSC)}, Asilomar, CA, Nov. 2005.

\bibitem{zsc.k06}
R.~Knopp, ``Two-way radio networks with a star topology,'' in \emph{Proc. int.
  Zurich Seminar on Communications}, Zurich, Feb. 2006.

\bibitem{ieeeIT.wnps10}
M.~P. Wilson, K.~Narayanan, H.~D. Pfister, and A.~Sprintson, ``Joint physical
  layer coding and network coding for bidirectional relaying,'' \emph{IEEE
  Trans. on Inform. Theory}, vol.~56, no.~11, pp. 5641--5654, Nov. 2010.

\bibitem{isit.rw06}
B.~Rankov and A.~Wittneben, ``Achievable rate regions for the two-way relay
  channel,'' in \emph{IEEE International Symposium on Information Theory},
  Seattle, Jul. 2006.

\bibitem{ieeeIT.cg79}
T.~M. Cover and A.~E. Gamal, ``Capacity theorems for the relay channel,''
  \emph{IEEE Trans. on Inform. Theory}, vol.~25, no.~5, pp. 572--584, Sep.
  1979.

\bibitem{acssc.sos07}
C.~Schnurr, T.~J. Oechtering, and S.~Stanczak, ``Achievable rates for the
  restricted half-duplex two-way relay channel,'' in \emph{41st Asilomar
  Conference on Signals, Systems and Computers (ACSSC)}, Asilomar, CA, Nov.
  2007.

\bibitem{sarnoff.kdmt08}
S.~J. Kim, N.~Devroye, P.~Mitran, and V.~Tarokh, ``Comparison of bi-directional
  relaying protocols,'' in \emph{IEEE Sarnoff Symposium}, Princeton, NJ, Apr.
  2008.

\bibitem{allerton.gtn08}
D.~Gunduz, E.~Tuncel, and J.~Nayak, ``Rate regions for the separated two-way
  relay channel,'' in \emph{46th Annual Allerton Conf. Comm. Control
  Computing}, Illinois, Sep. 2008, p. 1333–1340.

\bibitem{book.conway98}
J.~H. Conway and N.~J. Sloane, \emph{{Sphere packings, lattices and
  groups}}.\hskip 1em plus 0.5em minus 0.4em\relax New York, USA:
  Springe-Verlag, 3rd ed., 1998.

\bibitem{atc.skszd12}
S.~Smirani, M.~Kamoun, M.~Sarkiss, A.~Zaidi, and P.~Duhamel, ``Wyner-ziv
  lattice coding for two-way relay channel,'' in \emph{ATC 2012}, Hanoi,
  Vietnam, Oct. 2012.

\bibitem{ieeeIT.zse02}
R.~Zamir, S.~Shamai, and U.~Erez, ``Nested linear/lattice codes for structured
  multiterminal binning,'' \emph{IEEE Trans. on Inform. Theory}, vol.~48,
  no.~6, pp. 1250 -- 1276, Jun. 2002.

\bibitem{ieeeIT.ntg10}
J.~Nayak, E.~Tuncel, and D.~Gunduz, ``Wyner-ziv coding over broadcast channels:
  Digital schemes,'' \emph{IEEE Trans. on Inform. Theory}, vol.~56, no.~4, pp.
  1782--1799, Apr. 2010.

\bibitem{ieeeIT.ncl10}
W.~Nam, S.-Y. Chung, and Y.~H. Lee, ``Capacity of the gaussian two-way relay
  channel to within $\frac{1}{2}$ bit,'' \emph{IEEE Trans. on Inform. Theory},
  vol.~56, no.~11, pp. 5488--5494, Nov. 2010.

\bibitem{ieeeWC.twym12}
Y.~Tian, D.~Wu, C.~Yang, and A.~F. Molisch, ``Asymmetric two-way relay with
  doubly nested lattice codes,'' \emph{IEEE Trans. on Wireless Communications},
  vol.~11, no.~2, pp. 694--702, Feb. 2012.

\bibitem{ieeeIT.ez04}
U.~Erez and R.~Zamir, ``Achieving $\frac{1}{2} \log(1 + \text{SNR})$ on the
  {AWGN} channel with lattice encoding and decoding,'' \emph{IEEE Trans. on
  Inform. Theory}, vol.~50, no.~10, pp. 2293--2314, Oct. 2004.

\bibitem{ieeeIT.zf96}
R.~Zamir and M.~Feder, ``On lattice quantization noise,'' \emph{IEEE Trans. on
  Inform. Theory}, vol.~42, no.~4, pp. 1152--1159, Jul. 1996.

\bibitem{ieeeIT.poltyrev94}
G.~Poltyrev, ``On coding without restrictions for the {AWGN} channel,''
  \emph{IEEE Trans. on Inform. Theory}, vol.~40, no.~52, pp. 409--417, Mar.
  1994.

\bibitem{ieeeIT.elz05}
U.~Erez, S.~Litsyn, and R.~Zamir, ``Lattices which are good for (almost)
  everything,'' \emph{IEEE Trans. on Inform. Theory}, vol.~51, no.~10, pp.
  3401--3416, Oct. 2005.

\bibitem{book.cover91}
T.~Cover and J.~Thomas, \emph{Elements of Information Theory}.\hskip 1em plus
  0.5em minus 0.4em\relax New York: Wiley, 1991.

\bibitem{if.wz78}
A.~Wyner, ``The rate-distortion function for source coding with side
  information at the decoder-{II}: General sources,'' \emph{Information and
  Control}, vol.~38, no.~1, pp. 60--80, Jul. 1978.

\bibitem{ieeeIT.kdmt11}
S.~J. Kim, N.~Devroye, P.~Mitran, and V.~Tarokh, ``Achievable rate regions and
  performance comparison of half duplex bi-directional relaying protocols,''
  \emph{IEEE Trans. on Inform. Theory}, vol.~57, no.~10, pp. 6405--6418, 2011.

\end{thebibliography}
\end{document}